\documentclass[12pt,journal,onecolumn, draftcls]{IEEEtran}

\usepackage{epsfig}
\usepackage{times}
\usepackage{float}
\usepackage{afterpage}
\usepackage{amsmath}
\usepackage{amstext}
\usepackage{soul}
\usepackage{amssymb,bm}
\usepackage{latexsym}
\usepackage{color}
\usepackage{graphicx}
\usepackage{amsmath}
\usepackage{amsthm}
\usepackage{graphicx}
\usepackage[center]{caption}
\usepackage{subfig}
\usepackage{graphicx}
\usepackage{booktabs}
\usepackage{multicol}
\usepackage{lipsum}
\usepackage{dblfloatfix}
\usepackage{mathrsfs}
\usepackage{cite}
\usepackage{tikz}
\usepackage{pgfplots}
\pgfplotsset{compat=newest}

\allowdisplaybreaks

\usepackage{algorithm}
\usepackage{algpseudocode}
\algrenewcommand\algorithmicprocedure{\small \textbf{\textsf{procedure}}}
\algrenewtext{Procedure}[2]%
{\algorithmicprocedure\ \normalsize \textsc{\textrm{#1}}#2}
\algnewcommand\And{\textbf{and} }

\makeatletter
\newcommand{\removelatexerror}{\let\@latex@error\@gobble}
\def\NAT@spacechar{~}%
\makeatother

\newcommand{\bbC}{\mathbb{C}}
\newcommand{\rmd}{\mathrm{d}}
\newcommand{\bbE}{\mathbb{E}}\newcommand{\rme}{\mathrm{e}}

\newcommand{\bbI}{\mathbb{I}}

\newcommand{\bbN}{\mathbb{N}}\newcommand{\rmN}{\mathrm{N}}

\newcommand{\bbR}{\mathbb{R}}

\newcommand{\sfB}{\mathsf{B}}

\newcommand{\sfD}{\mathsf{D}}

\newcommand{\bfI}{\mathbf{I}}\newcommand{\sfI}{\mathsf{I}}

\newcommand{\bfN}{\mathbf{N}}

\newcommand{\bfp}{\mathbf{p}}

\newcommand{\sfR}{\mathsf{R}}

\newcommand{\bfW}{\mathbf{W}}
\newcommand{\bfX}{\mathbf{X}}\newcommand{\bfx}{\mathbf{x}}
\newcommand{\bfY}{\mathbf{Y}}\newcommand{\bfy}{\mathbf{y}}
\newcommand{\bfZ}{\mathbf{Z}}

\newcommand{\cB}{\mathcal{B}}
\newcommand{\cC}{\mathcal{C}}
\newcommand{\cD}{\mathcal{D}}

\newcommand{\sfh}{\mathsf{h}}
\newcommand{\cI}{\mathcal{I}}

\newcommand{\scrN}{\mathscr{N}}

\newcommand{\scrS}{\mathscr{S}}

\newcommand{\supp}{{\mathsf{supp}}}
\newcommand{\bfrho}{\boldsymbol{\rho}}

\theoremstyle{mystyle}
\newtheorem{theorem}{Theorem}
\theoremstyle{mystyle}
\newtheorem{lemma}{Lemma}
\theoremstyle{mystyle}
\theoremstyle{mystyle}
\theoremstyle{mystyle}
\newtheorem{definition}{Definition}
\theoremstyle{remark}
\newtheorem{rem}{Remark}
\theoremstyle{mystyle}
\theoremstyle{mystyle}
\theoremstyle{mystyle}
\theoremstyle{discussion}
\theoremstyle{mystyle}
\theoremstyle{mystyle}

 \usepackage{enumitem}

\begin{document}
\bstctlcite{IEEEexample:BSTcontrol}

\title{Amplitude Constrained Vector Gaussian Wiretap Channel: Properties of the Secrecy-Capacity-Achieving Input Distribution}

\author{\thanks{{Part of this work was presented at the 2021 IEEE Information
		Theory Workshop~\cite{barletta2021scalar} and at the 2022 IEEE International Symposium on Information Theory~\cite{favano2022capacity}.}}
Antonino Favano\thanks{A. Favano  is with the Dipartimento di Elettronica, Informazione e Bioingegneria, Politecnico di Milano, Milano, 20133, Italy, and with the Consiglio Nazionale delle Ricerche, Milano, 20133, Italy. (e-mail: antonino.favano@polimi.it).}, Luca Barletta\thanks{L. Barletta  is with the Dipartimento di Elettronica, Informazione e Bioingegneria, Politecnico di Milano, Milano, 20133, Italy. (e-mail: luca.barletta@polimi.it).}, and  Alex Dytso\thanks{A. Dytso is with the Department of Electrical and Computer Engineering, New Jersey Institute of Technology, Newark,  NJ 07102, USA   (e-mail: alex.dytso@njit.edu).
}}
\maketitle
\begin{abstract}
This paper studies secrecy-capacity of an $n$-dimensional Gaussian wiretap channel under a peak-power constraint. This work determines the largest peak-power constraint $\bar{\sfR}_n$ such that an input distribution uniformly distributed on a single sphere is optimal; this regime is termed the low amplitude regime. The asymptotic of $\bar{\sfR}_n$ as $n$ goes to infinity is completely characterized as a function of noise variance at both receivers. Moreover, the secrecy-capacity is also characterized in a form amenable for computation. Several numerical examples are provided, such as the example of the secrecy-capacity-achieving distribution beyond the low amplitude regime. Furthermore, for the scalar case $(n=1)$ we show that the secrecy-capacity-achieving input distribution is discrete with finitely many points at most of the order of $\frac{\sfR^2}{\sigma_1^2}$, where $\sigma_1^2$ is the variance of the Gaussian noise over the legitimate channel.
\end{abstract}

\section{Introduction} 
Consider the vector Gaussian wiretap channel with outputs 
\begin{align}
\bfY_1&= \bfX+\bfN_1,\\
\bfY_2&=\bfX+\bfN_2,
\end{align}
where $\bfX \in \bbR^n$ and where 
$\bfN_1 \sim \mathcal{N}(\mathbf{0}_n,\sigma_1^2 \bfI_n)$ and $\bfN_2 \sim \mathcal{N}(\mathbf{0}_n,\sigma_2^2 \bfI_n)$, and with $(\bfX,\bfN_1,\bfN_2)$ mutually independent. The output $\bfY_1$ is observed by the legitimate receiver whereas the output $\bfY_2$ is observed by the malicious receiver. In this work, we are interested in the scenario where the input $\bfX$ is limited by a peak-power constraint or amplitude constraint and assume that $\bfX \in \cB_0(\sfR) = \{ \bfx : \: \| \bfx \| \leq \sfR \}$, i.e., $  \cB_0(\sfR)$ is an $n$-ball centered at ${\bf 0}$ of radius $\sfR$.
For this setting, the secrecy-capacity is given by 
\begin{align}
C_s(\sigma_1^2, \sigma_2^2, \sfR, n) &= \max_{\bfX \in  \cB_0(\sfR) }  I(\bfX; \bfY_1) - I(\bfX; \bfY_2) \\
&= \max_{\bfX \in  \cB_0(\sfR)}   I(\bfX; \bfY_1 | \bfY_2), \label{eq:Secracy_CAP}
\end{align}
where the last expression holds due to the degraded nature of the channel. 
It can be shown that for $\sigma_1^2 \ge \sigma^2_2$ the secrecy-capacity is equal to zero. Therefore, in the remaining, we assume that $\sigma_1^2 < \sigma^2_2$.

We are interested in studying the input distribution $P_{\bfX^\star}$ that maximizes \eqref{eq:Secracy_CAP} in the low (but not vanishing) amplitude regime. Since closed-form expressions for secrecy-capacity are rare, we derive the secrecy-capacity in an integral form that is easy to evaluate. For the scalar case $(n=1)$ we establish an upper bound on the number of mass points of $P_{X^\star}$, valid for any amplitude regime. We also argue in Section~\ref{sec:Connection_Other_Problem} that the solution to the secrecy-capacity can shed light on other problems seemingly unrelated to security.  The paper also provides a number of numerical simulations of $P_{\bfX^\star}$ and $C_s$, the data for which are made available at~\cite{GithubData}.

  \subsection{Literature Review} 
The wiretap channel was introduced by Wyner in \cite{wyner1975wire}, who also established the secrecy-capacity of the degraded wiretap channel. The results of~\cite{wyner1975wire} were extended to the Gaussian wiretap channel in~\cite{GaussianWireTap}. The wiretap channel plays a central role in network information theory; the interested reader is referred to \cite{bloch2011physical,Oggier2015Wiretap,Liang2009Security,poor2017wireless,mukherjee2014principles} and reference therein for a detailed treatment of the topic. Furthermore, for an in-depth discussion on the wiretap fading channel refer to~\cite{gopala2008secrecy,bloch2008wireless,khisti2008secure,liang2008secure}.

In~\cite{GaussianWireTap} it was shown that the secrecy-capacity-achieving input distribution of the Gaussian wiretap channel, under an average-power constraint, is Gaussian. In~\cite{shafiee2009towards}, the authors investigated the Gaussian wiretap channel consisting of two antennas both at the transmitter and receiver side and of a single antenna for the eavesdropper. The secrecy-capacity of the MIMO wiretap channel was characterized in \cite{khisti2010secure} and \cite{oggier2011secrecy} where the Gaussian input was shown to be optimal. An elegant proof, using the I-MMSE relationship \cite{I-MMSE}, of optimality of Gaussian input, was given in \cite{bustin2009mmse}. Moreover, an alternative approach in the characterization of the secrecy-capacity of a MIMO wiretap channel was proposed in~\cite{liu2009note}. In~\cite{loyka2015algorithm} and~\cite{loyka2016optimal} the authors discuss the optimal signaling for secrecy rate maximization under average power constraint.

The secrecy-capacity of the Gaussian wiretap channel under the peak-power constraint has received far less attention. The secrecy-capacity of the scalar Gaussian wiretap channel with an amplitude and power constraint was considered in \cite{ozel2015gaussian} where the authors showed that the capacity-achieving input distribution $P_{X^\star}$ is discrete with finitely many support points.

The work of~\cite{ozel2015gaussian} was extended to noise-dependent channels by Soltani and Rezki in~\cite{soltani2018optical}. For further studies on the properties of the secrecy-capacity-achieving input distribution for a class of degraded wiretap channels, refer to~\cite{soltani2021degraded,nam2019secrecy,DytsoITWwiretap2018}.

The secrecy-capacity for the vector wiretap channel with a peak-power constraint was considered in \cite{DytsoITWwiretap2018} where it was shown that the optimal input distribution is concentrated on finitely many co-centric shells. 
 
\subsection{Contributions and Paper Outline}
In Section~\ref{sec:Assump&Motiv} we introduce mathematical tools, assumptions and definitions used throughout the paper. Specifically, in Section~\ref{sec:small_amp_regime} we give a definition of low amplitude regime. Moreover, in Section~\ref{sec:Connection_Other_Problem} we show how the wiretap channel can be seen as a generalization of point-to-point channels and the evaluation of the largest minimum mean square error (MMSE), both under the assumption of amplitude constrained input.

In Section~\ref{sec:main_results} we detail our main results. Theorem~\ref{thm:Char_Small_Amplitude} defines the radius $\bar{\sfR}_n$ below which we are in the low amplitude regime, i.e., the optimal input distribution is composed of a single shell.  Theorem~\ref{thm:large_n_beh} characterizes the asymptotic behavior of $\bar{\sfR}_n$ as $n$ goes to infinity. Furthermore, Theorem \ref{thm:Main_Results_Scalar} gives an implicit and an explicit upper bound on the number of mass points of the secrecy-capacity-achieving input distribution when $n=1$.

In Section~\ref{sec:Cs_small_amp_regime} we derive the secrecy-capacity expression for the low amplitude regime in Theorem~\ref{thm:Capacitiy_Small}. We also investigate its behavior when the number of antennas $n$ goes to infinity. 

Section~\ref{sec:beyond_small_amp_regime} extends the investigation of the secrecy-capacity beyond the low amplitude regime. We numerically estimate both the optimal input pmf and the resulting capacity via an algorithmic procedure based on the KKT conditions introduced in Lemma~\ref{lem:KKT}.

Section~\ref{sec:thm:Char_Small_Amplitude}, Section~\ref{sec:large_n_beh}, Section~\ref{Sec:main_result_scalar} and Section~\ref{sec:thm:Capacitiy_Small} provide the proof for Theorem~\ref{thm:Char_Small_Amplitude}, Theorem~\ref{thm:large_n_beh}, Theorem~\ref{thm:Main_Results_Scalar} and Theorem~\ref{thm:Capacitiy_Small}, respectively.

Finally, Section~\ref{sec:conclusion} concludes the paper.

\subsection{Notation}
We use bold letters for vectors ($\bfx$) and uppercase letters for random variables ($X$). We denote by $\| \bfx \|$ the Euclidean norm of the vector $\bfx$.
Given a random variable $X$, its probability density function (pdf), mass function (pmf), and cumulative distribution function are denoted by $f_X$, $P_X$, and $F_X$, respectively. The support set of $P_\bfX$ is denoted and defined as
\begin{align}
	\supp(P_{\bfX})&=\{\bfx: \text{ for every open set $ \mathcal{D} \ni \bfx $ } \notag\\ 
	&\quad \qquad \text{
		we have that $P_{\bfX}( \mathcal{D})>0$} \}. 
\end{align} 
 We denote by $\mathcal{N}(\boldsymbol{\mu},\mathsf{\Sigma})$ a multivariate Gaussian distribution with mean vector $\boldsymbol{\mu}$ and covariance matrix $\mathsf{\Sigma}$. The pdf of a Gaussian random variable with zero mean and variance $\sigma^2$ is denoted by  $\phi_{\sigma}( \cdot)$. We denote by $\chi^2_{n}(\lambda)$ the noncentral chi-square distribution with $n$ degrees of freedom and with noncentrality parameter $\lambda$. We represent the $n \times 1$ vector of zeros by $\mathbf{0}_n$ and the $n \times n$ identity matrix by $\bfI_n$.
 Furthermore, we represent by $\sfD$ the relative entropy.
The minimum mean squared error is denoted by 
\begin{align}
	{\rm mmse}(\bfX| \bfX+\bfN)= \bbE \left[ \| \bfX-\bbE[\bfX| \bfX+\bfN] \|^2 \right].
\end{align}
The modified Bessel function of the first kind of order $v \ge 0 $ will be denoted by $\sfI_v(x), x\in \bbR$. The following ratio of the Bessel functions will be commonly used in this work:
\begin{equation}
	\sfh_v(x) =\frac{\sfI_v(x)}{\sfI_{v-1}(x)},\, x\in \bbR,\, v\ge 0. 
\end{equation}
Finally, the number of zeros (counted in accordance with their multiplicities) of a function $f \colon \mathbb{R} \to \mathbb{R} $  on the interval $\cI$ is denoted by  $\rmN(\cI, f)$. Similarly, if $f  \colon \bbC \to \bbC$ is a function on the complex domain, $\rmN(\cD, f)$ denotes the number of its  zeros within the region $\cD$.

\section{Preliminaries } \label{sec:Assump&Motiv}

\subsection{Oscillation Theorem}\label{sec:oscillation}
In this work, we will often need to upper bound the number of oscillations of a function, \emph{i.e.}, its number of sign changes. This is useful, for example, to bound the number of zeros of a function, or the number of roots of an equation. To be more precise, let us define the number of sign changes as follows.
\begin{definition}[Sign Changes of a Function]  The number of sign changes of a function $\xi: \Omega \to \mathbb{R}$ is given by 
	\begin{equation}
	\scrS(\xi) = \sup_{m\in \bbN } \left\{\sup_{y_1< \cdots< y_m  \subseteq \Omega} \scrN \{ \xi (y_i) \}_{i=1}^m\right\} \text{,}
	\end{equation}
	where  $\scrN\{ \xi (y_i) \}_{i=1}^m$ is the number of sign changes of the sequence $\{ \xi (y_i) \}_{i=1}^m $.
\end{definition}

In~\cite{karlin1957polya}, Karlin noticed that some integral transformations have a \emph{variation-diminishing} property, which is described in the following theorem.
\begin{theorem}[Oscillation Theorem]\label{thm:OscillationThoerem} Given domains $\bbI_1 $ and $\bbI_2$, let $p\colon \bbI_1\times \bbI_2  \to \bbR$ be a strictly totally positive
	kernel.\footnote{A function $f:\bbI_1 \times \bbI_2 \to \bbR$ is said to be a totally positive kernel of order $n$ if $\det\left([f(x_i,y_j)]_{i,j = 1}^{m}\right) >0 $ for all $1\le m \le n $, and for all $x_1< \cdots < x_m \in \bbI_1  $, and $y_1< \cdots < y_m \in \bbI_2$. If $f$ is  totally positive kernel of order $n$ for all $n\in \bbN$, then $f$ is a strictly totally positive kernel.} For an arbitrary $y$, suppose $p(\cdot, y)\colon \bbI_1 \to \bbR $ is an $n$-times differentiable function. Assume that $\mu$ is a measure on $\bbI_2 $, and let $\xi \colon \bbI_2 \to \bbR $ be a function with $\scrS(\xi) = n$. For $x\in \bbI_1$, define
	\begin{equation}
	\Xi(x)=  \int  \xi (y) p(x ,y) {\rm d} \mu(y) \text{.} \label{eq:Integral_Transform}
	\end{equation}
	If $\Xi \colon \bbI_1 \to \bbR$ is an $n$-times differentiable function, then either $\rmN(\bbI_1, \Xi) \le n$, or $\Xi\equiv 0$.  
\end{theorem} 
The above theorem says that the number of zeros of a function $\Xi$, which is the output of the integral transformation, is less than the number of sign changes of the function $  \xi  $, which is the input to the integral transformation.

\subsection{Assumptions}
\label{sec:Assumptions}
Consider the following function: for $y \in \mathbb{R}^+$
\begin{align}
&G_{\sigma_1,\sigma_2,\sfR,n}(y)\notag\\
&=\frac{\bbE\left[\frac{\sfR}{\|y+\bfW\|}\sfh_{\frac{n}{2}}\left(\frac{\sfR}{\sigma_2^2}\| y+\bfW\|\right)-1  \right]}{\sigma_2^2} -\frac{\frac{\sfR}{y}\sfh_{\frac{n}{2}}\left(\frac{\sfR}{\sigma_1^2}y\right) -1 }{\sigma_1^2},  \label{eq:Definition_of_G_function}
	\end{align} 
	where $\bfW \sim  {\cal N}(\mathbf{0}_{n+2},(\sigma_2^2-\sigma_1^2)\bfI_{n+2})$. Notice that the function $G_{\sigma_1,\sigma_2,\sfR,n}$ is related to the derivative of the secrecy-density. (See the proof of Theorem~\ref{thm:equivalent_condition}.)
	
	In this work, in order to make progress on the secrecy-capacity, we make the following \emph{conjecture} about the ratio of the Bessel functions: for all $\sfR \ge 0, \sigma_2 \ge \sigma_1 \ge 0$ and $n \in \mathbb{N}$, the function $y \mapsto G_{\sigma_1,\sigma_2,\sfR,n}(y)$ has \emph{at most} one sign change.   	
	In general, proving that $G_{\sigma_1,\sigma_2,\sfR,n}$ has at most one sign change is not easy. 	However, extensive numerical evaluations show that this property holds for any $n, \sfR, \sigma_1, \sigma_2$; see Appendix~\ref{app:Examples_G_func} for the examples.

	  Therefore, the problem boils down to showing that there is at most one sign change for $y>0$. 
Using this, we can give a sufficient condition for this conjecture to be true. Note that
\begin{align}
G_{\sigma_1,\sigma_2,\sfR,n}(y)&\ge-\frac{1}{\sigma_2^2}+\frac{1}{\sigma_1^2}-\frac{\sfR}{\sigma_1^2 y}\sfh_{\frac{n}{2}}\left(\frac{\sfR}{\sigma_1^2}y\right) \label{eq:LB_on_h} \\
&\ge -\frac{1}{\sigma_2^2}+\frac{1}{\sigma_1^2}-\frac{\sfR^2}{\sigma_1^4 n}, \label{eq:UB_on_h}
\end{align}
which is nonnegative, hence has no sign change, for 
\begin{equation}
\sfR < \sigma_1^2 \sqrt{n \left(\frac{1}{\sigma_1^2}-\frac{1}{\sigma_2^2}\right)},
\end{equation}
for all $y\ge 0$. The inequality in~\eqref{eq:LB_on_h} follows from $\sfh_{\frac{n}{2}}(x)\ge 0$ for $x\ge 0$; and~\eqref{eq:UB_on_h} follows from $\sfh_{\frac{n}{2}}(x)\le \frac{x}{n}$ for $x\ge 0$ and $n\in \mathbb{N}$.

\subsection{Low Amplitude Regime} \label{sec:small_amp_regime}

In this work, a low amplitude regime is defined as follows. 
\begin{definition} Let $\bfX_{\sfR} \sim P_{\bfX_{\sfR}}$ be uniform on $\cC(\sfR)=\{ \bfx :  \|\bfx\|=\sfR \}$. The capacity in \eqref{eq:Secracy_CAP} is said to be in the low amplitude regime if $\sfR \le \bar{\sfR}_n(\sigma_1^2,\sigma_2^2)$ where
\begin{equation}
\bar{\sfR}_n(\sigma_1^2,\sigma_2^2)= \max \left\{ \sfR:  P_{\bfX_{\sfR}} =\arg  \max_{\bfX \in  \cB_0(\sfR)}   I(\bfX; \bfY_1 | \bfY_2)  \right \}. \label{eq:small_amplitude_def}
\end{equation}
If the set in \eqref{eq:small_amplitude_def} is empty, then we assign $\bar{\sfR}_n(\sigma_1^2,\sigma_2^2)=0$. 
\end{definition}
The quantity $\bar{\sfR}_n(\sigma_1^2,\sigma_2^2)$ represents the largest radius $\sfR$ for which $P_{\bfX_{\sfR}}$ is secrecy-capacity-achieving. 

One of the main objectives of this work is to characterize $\bar{\sfR}_n(\sigma_1^2,\sigma_2^2)$. 
\subsection{Connections to Other Optimization Problems} 
\label{sec:Connection_Other_Problem}

The distribution $ P_{\bfX_{\sfR}}$ occurs in a variety of statistical and information-theoretic applications. For example, consider the following two optimization problems:
\begin{align}
 \max_{\bfX \in  \cB_0(\sfR)}&   I(\bfX; \bfX+\bfN),\\
  \max_{\bfX \in  \cB_0(\sfR)}& {\rm mmse}(\bfX| \bfX+\bfN),
 \end{align} 
where $\bfN \sim \mathcal{N}(\mathbf{0}_n,\sigma^2 \bfI_n)$. The first problem seeks to characterize the capacity of the point-to-point channel under an amplitude constraint,
 and the second problem seeks to find the largest minimum mean squared error under the assumption that the signal has bounded amplitude; the interested reader is referred to \cite{dytsoMI_est_2019,favano2021capacity,berry1990minimax} for a detailed background on both problems.  
 
Similarly to the wiretap channel, we can define the low amplitude regime for both problems as the largest $\sfR$ such that $ P_{\bfX_{\sfR}}$ is optimal and denote these by $\bar{\sfR}_n^\text{ptp}(\sigma^2)$ and $\bar{\sfR}_n^\text{MMSE}(\sigma^2)$.
We now argue that both $\bar{\sfR}_n^\text{ptp}(\sigma^2)$ and $\bar{\sfR}_n^\text{MMSE}(\sigma^2)$
 can be seen as a special case of the wiretap solution. Hence, the wiretap channel provides and interesting unification and generalization of these two problems.  
 
 First, note that the point-to-point solution can be recovered from the wiretap by simply specializing the wiretap channel to the point-to-point channel, that is 
 \begin{align} \label{eq:Rptp}
 \bar{\sfR}_n^\text{ptp}(\sigma^2)= \lim_{\sigma_2 \to \infty} \bar{\sfR}_n(\sigma^2,\sigma_2^2).
 \end{align} 
 Second, to see that the MMSE solution can be recovered from the wiretap recall that by the I-MMSE relationship \cite{I-MMSE}, we have that 
 \begin{align}
 & \max_{\bfX \in  \cB_0(\sfR) }  I(\bfX; \bfY_1) - I(\bfX; \bfY_2) \notag \\
  &=  \max_{\bfX \in  \cB_0(\sfR) }  \frac{1}{2} \int_{\sigma_1^2}^{\sigma_2^2} \frac{ {\rm mmse}(\bfX| \bfX+ \sqrt{s}\bfZ)}{s^2 } \rmd s
 \end{align} 
 where $\bfZ$ is standard Gaussian.  Now note that if we choose $\sigma_2^2=\sigma_1^2+\epsilon$ for some small enough $\epsilon>0$, we arrive at
 \begin{align}
 & \max_{\bfX \in  \cB_0(\sfR) }  I(\bfX; \bfY_1) - I(\bfX; \bfY_2) \\
 &=    \max_{\bfX \in  \cB_0(\sfR) }   \frac{\epsilon}{2}  \frac{ {\rm mmse}(\bfX| \bfX+ \sqrt{\sigma_1^2}\bfZ)}{\sigma_1^4 }.
 \end{align}  
 Consequently, for a small enough $\epsilon>0$, 
 \begin{equation}\label{eq:reduction_to_mmse}
 \bar{\sfR}_n^\text{MMSE}(\sigma^2)=  \bar{\sfR}_n(\sigma^2,\sigma^2+\epsilon).
 \end{equation} 

\section{Main Results} \label{sec:main_results}

\subsection{Characterizing the Low Amplitude Regime} 

Our first main result characterizes the low amplitude regime.  
\begin{theorem}\label{thm:Char_Small_Amplitude}
Consider a function 
\begin{align}
 f(\sfR)
&=\int_{\sigma_1^2}^{\sigma_2^2} \frac{\bbE \left[      \mathsf{h}_{\frac{n}{2}}^2\left(  \frac{\|  \sqrt{s}\bfZ\| \sfR}{s} \right) +     \mathsf{h}_{\frac{n}{2}}^2\left(  \frac{\|  \sfR+\sqrt{s}\bfZ\| \sfR}{s} \right) \right]-1}{s^2} \rmd s
\end{align} 
where $\bfZ \sim {\cal N}(\mathbf{0}_n,\bfI_n)$.
The input $\bfX_{\sfR}$ is secrecy-capacity-achieving if  and only if $\sfR \le \bar{\sfR}_n(\sigma_1^2,\sigma_2^2)$ where $\bar{\sfR}_n(\sigma_1^2,\sigma_2^2)$ is given as the solution of 
\begin{equation}
f(\sfR)=0.  \label{eq:Condition_for_optimality}
\end{equation} 

\end{theorem} 

\begin{rem} Note that \eqref{eq:Condition_for_optimality} always has a solution. To see this, observe that $f(0)=\frac{1}{\sigma_2^2}-\frac{1}{\sigma_1^2}<0$, and $f(\infty)=\frac{1}{\sigma_1^2}-\frac{1}{\sigma_2^2}>0$. Moreover, the solution is unique, because $f(\sfR)$ is monotonically increasing for $\sfR\ge 0$.
\end{rem}

The solution to \eqref{eq:Condition_for_optimality} needs to be found numerically.\footnote{To avoid any loss of accuracy in the numerical evaluation of $\sfh_v(x)$ for large values of $x$, we used the exponential scaling provided in the MATLAB implementation of $\sfI_v(x)$.} Since evaluating $f(\sfR)$ is rather straightforward and not time-consuming, we opted for a binary search algorithm.
\begin{table}[t]
	\caption{Values of $\bar{\sfR}^{\text{MMSE}}_n(1)$, $\bar{\sfR}_n(1,\sigma_2^2)$, and $\bar{\sfR}^{\text{ptp}}_n(1)$}
	\[\begin{tabular}{l c c c c c c}
		\toprule
		& \text{MMSE} & \multicolumn{4}{c@{}}{$\sigma_2^2$} & \text{ptp} \\
		\cmidrule(l){3-6} 
		$n$ &  & 1.001  & 1.5    & 10  & 1000    &  \\
		\midrule
		
1	&	1.057	&	1.057	&	1.161	&	1.518	&	1.664	&	1.666	\\
2	&	1.535	&	1.535	&	1.687	&	2.221	&	2.450	&	2.454	\\
3	&	1.908	&	1.909	&	2.098	&	2.768	&	3.061	&	3.065	\\
4	&	2.223	&	2.224	&	2.444	&	3.229	&	3.575	&	3.580	\\
5	&	2.501	&	2.501	&	2.750	&	3.634	&	4.026	&	4.031	\\
6	&	2.751	&	2.752	&	3.025	&	3.999	&	4.432	&	4.438	\\
7	&	2.981	&	2.982	&	3.278	&	4.334	&	4.805	&	4.811	\\
8	&	3.195	&	3.196	&	3.513	&	4.646	&	5.151	&	5.158	\\
9	&	3.395	&	3.396	&	3.733	&	4.937	&	5.475	&	5.483	\\
10	&	3.585	&	3.586	&	3.941	&	5.213	&	5.781	&	5.789	\\
11	&	3.765	&	3.766	&	4.139	&	5.475	&	6.072	&	6.080	\\
12	&	3.936	&	3.938	&	4.328	&	5.725	&	6.350	&	6.359	\\
13	&	4.101	&	4.102	&	4.509	&	5.964	&	6.616	&	6.625	\\
14	&	4.259	&	4.260	&	4.683	&	6.195	&	6.872	&	6.881	\\
15	&	4.412	&	4.413	&	4.851	&	6.417	&	7.119	&	7.128	\\
16	&	4.560	&	4.561	&	5.013	&	6.632	&	7.357	&	7.367	\\
17	&	4.702	&	4.704	&	5.170	&	6.839	&	7.588	&	7.598	\\
18	&	4.841	&	4.842	&	5.323	&	7.041	&	7.812	&	7.823	\\
19	&	4.976	&	4.977	&	5.471	&	7.238	&	8.030	&	8.041	\\
20	&	5.107	&	5.109	&	5.616	&	7.429	&	8.242	&	8.254	\\
21	&	5.235	&	5.237	&	5.756	&	7.615	&	8.449	&	8.461	\\
22	&	5.360	&	5.362	&	5.894	&	7.797	&	8.651	&	8.663	\\
23	&	5.483	&	5.484	&	6.028	&	7.974	&	8.848	&	8.860	\\
24	&	5.602	&	5.603	&	6.159	&	8.148	&	9.041	&	9.054	\\
25	&	5.719	&	5.720	&	6.288	&	8.318	&	9.230	&	9.243	\\
26	&	5.834	&	5.835	&	6.414	&	8.485	&	9.416	&	9.428	\\
27	&	5.946	&	5.948	&	6.538	&	8.649	&	9.597	&	9.610	\\
28	&	6.056	&	6.058	&	6.659	&	8.809	&	9.775	&	9.789	\\
29	&	6.165	&	6.166	&	6.778	&	8.967	&	9.951	&	9.964	\\
30	&	6.271	&	6.273	&	6.895	&	9.122	&	10.123	&	10.136	\\
31	&	6.376	&	6.378	&	7.010	&	9.274	&	10.292	&	10.306	\\
32	&	6.479	&	6.481	&	7.124	&	9.424	&	10.458	&	10.472	\\
33	&	6.580	&	6.582	&	7.235	&	9.571	&	10.622	&	10.636	\\
34	&	6.680	&	6.682	&	7.345	&	9.717	&	10.783	&	10.798	\\
35	&	6.779	&	6.780	&	7.453	&	9.860	&	10.942	&	10.957	\\
		
		\bottomrule
	\end{tabular}\]
\label{Table1}
\end{table}
 In Table~\ref{Table1}, we show the values of $\bar{\sfR}_n ( 1,\sigma_2^2 )$ for some values of~$\sigma_2^2$ and~$n$. Moreover, we report the values of $\bar{\sfR}_n^{\text{ptp}}(1)$ and $\bar{\sfR}_n^{\text{MMSE}}(1)$ from~\cite{dytsoMI_est_2019} in the first and the last row, respectively. As predicted by~\eqref{eq:Rptp}, we can appreciate the close match of the $\bar{\sfR}_n^{\text{ptp}}(1)$ row with the one of $\bar{\sfR}_n(1,1000)$. Similarly, the agreement between the $\bar{\sfR}_n^{\text{MMSE}}(1)$ row and the $\bar{\sfR}_n(1,1.001)$ row is justified by~\eqref{eq:reduction_to_mmse}.

\subsection{Large $n$ Asymptotics} 

We now use the result in Theorem~\ref{thm:Char_Small_Amplitude} to characterize the asymptotic behavior of $\bar{\sfR}_n(\sigma_1^2,\sigma_2^2)$.  In particular, it is shown that $\bar{\sfR}_n(\sigma_1^2,\sigma_2^2)$ increases as $\sqrt{n}$. 
\begin{theorem}\label{thm:large_n_beh}  For $\sigma_1^2 \le \sigma^2_2$  
\begin{equation} \label{eq:c_asym}
\lim_{n \to \infty} \frac{\bar{\sfR}_n(\sigma_1^2,\sigma_2^2)}{\sqrt{n}}=c(\sigma_1^2,\sigma_2^2),
\end{equation}
where $c(\sigma_1^2,\sigma_2^2)$ is the solution of 
\begin{equation}
 \int_{\sigma_1^2}^{\sigma_2^2} \frac{{ \frac{c^2 }{ \left(  \frac{\sqrt{s}}{2}+\sqrt{ \frac{s}{4} + c^2} \right)^2}} +      \frac{ c^2 (c^2+ s)}{ \left( \frac{s}{2}+\sqrt{ \frac{s^2}{4} +c^2( c^2+ s)  } \right)^2} -1}{s^2} \rmd s =0.
\end{equation} 
\end{theorem} 
\begin{IEEEproof}
See Section~\ref{sec:large_n_beh}. 
\end{IEEEproof}%
\begin{figure}[t]
	\centering
%
%
\begin{tikzpicture}

\begin{axis}[%
width=\linewidth,
height=0.8\linewidth,
xmin=1,
xmax=35,
xlabel style={font=\color{white!15!black}},
xlabel={$n$},
ymin=1,
ymax=2,
ylabel style={font=\color{white!15!black}},
axis background/.style={fill=white},
xmajorgrids,
ymajorgrids,
legend style={legend cell align=left, align=left, draw=white!15!black,at={(0.97,0.45)},anchor=east}
]
\addplot [color=red, line width=0.8pt]
  table[row sep=crcr]{%
1	1.15125151780705\\
2	1.15125151780705\\
3	1.15125151780705\\
4	1.15125151780705\\
5	1.15125151780705\\
6	1.15125151780705\\
7	1.15125151780705\\
8	1.15125151780705\\
9	1.15125151780705\\
10	1.15125151780705\\
11	1.15125151780705\\
12	1.15125151780705\\
13	1.15125151780705\\
14	1.15125151780705\\
15	1.15125151780705\\
16	1.15125151780705\\
17	1.15125151780705\\
18	1.15125151780705\\
19	1.15125151780705\\
20	1.15125151780705\\
21	1.15125151780705\\
22	1.15125151780705\\
23	1.15125151780705\\
24	1.15125151780705\\
25	1.15125151780705\\
26	1.15125151780705\\
27	1.15125151780705\\
28	1.15125151780705\\
29	1.15125151780705\\
30	1.15125151780705\\
31	1.15125151780705\\
32	1.15125151780705\\
33	1.15125151780705\\
34	1.15125151780705\\
35	1.15125151780705\\
};
\addlegendentry{\scriptsize $c(1,\sigma_2^2) \ \eqref{eq:c_asym}$}

\addplot [color=black, line width=0.8pt, mark=*, mark size =1pt, mark options={solid, black}]
  table[row sep=crcr]{%
1	1.05700754961514\\
2	1.08567631436701\\
3	1.10184786913882\\
4	1.11190914815219\\
5	1.11867291122447\\
6	1.12349710272807\\
7	1.12709773157121\\
8	1.12988199389348\\
9	1.13209646405149\\
10	1.13389840141803\\
11	1.13539249136889\\
12	1.13665098589732\\
13	1.1377252939573\\
14	1.13865288306867\\
15	1.13946185044196\\
16	1.14017348091489\\
17	1.14080429167319\\
18	1.14136728158719\\
19	1.14187282962213\\
20	1.1423292467616\\
21	1.1427433827209\\
22	1.14312083565491\\
23	1.14346628037165\\
24	1.14378358654872\\
25	1.14407607154453\\
26	1.14434654435366\\
27	1.14459736408627\\
28	1.1448306274395\\
29	1.14504810200458\\
30	1.14525134228772\\
31	1.14544169089822\\
32	1.14562035265051\\
33	1.14578835697793\\
34	1.14594661770954\\
35	1.14609598552094\\
};
\addlegendentry{\scriptsize $\frac{\Bar{\mathsf{R}}_n(1,\sigma_2^2)}{\sqrt{n}}$}

\node[below right, align=left]
at (axis cs:15,1.147) {\scriptsize $\sigma_2^2 = 1.001$};
\addplot [color=red, line width=0.8pt, forget plot]
  table[row sep=crcr]{%
1	1.26546217419275\\
2	1.26546217419275\\
3	1.26546217419275\\
4	1.26546217419275\\
5	1.26546217419275\\
6	1.26546217419275\\
7	1.26546217419275\\
8	1.26546217419275\\
9	1.26546217419275\\
10	1.26546217419275\\
11	1.26546217419275\\
12	1.26546217419275\\
13	1.26546217419275\\
14	1.26546217419275\\
15	1.26546217419275\\
16	1.26546217419275\\
17	1.26546217419275\\
18	1.26546217419275\\
19	1.26546217419275\\
20	1.26546217419275\\
21	1.26546217419275\\
22	1.26546217419275\\
23	1.26546217419275\\
24	1.26546217419275\\
25	1.26546217419275\\
26	1.26546217419275\\
27	1.26546217419275\\
28	1.26546217419275\\
29	1.26546217419275\\
30	1.26546217419275\\
31	1.26546217419275\\
32	1.26546217419275\\
33	1.26546217419275\\
34	1.26546217419275\\
35	1.26546217419275\\
};
\addplot [color=black, line width=0.8pt, mark=*, mark size =1pt, mark options={solid, black}, forget plot]
  table[row sep=crcr]{%
1	1.16127361761653\\
2	1.19315741430016\\
3	1.21106514355155\\
4	1.22217627600009\\
5	1.2296336837499\\
6	1.23494726245077\\
7	1.23891067505508\\
8	1.2419741272445\\
9	1.24440997324439\\
10	1.24639161344061\\
11	1.24803444050257\\
12	1.24941806043064\\
13	1.25059902990421\\
14	1.25161869229161\\
15	1.2525078699636\\
16	1.25329001221689\\
17	1.25398330182204\\
18	1.25460204908811\\
19	1.25515761954344\\
20	1.25565921147753\\
21	1.2561143261815\\
22	1.25652909690446\\
23	1.25690868414368\\
24	1.25725736984404\\
25	1.25757874778852\\
26	1.25787596143686\\
27	1.25815156205166\\
28	1.25840788367657\\
29	1.25864684574435\\
30	1.25887015119169\\
31	1.25907929901797\\
32	1.2592755979502\\
33	1.2594601888743\\
34	1.25963408253852\\
35	1.2597982055379\\
};
\node[above right, align=left]
at (axis cs:15,1.262) {\scriptsize $\sigma_2^2 = 1.5$};
\addplot [color=red, line width=0.8pt, forget plot]
  table[row sep=crcr]{%
1	1.67419101387548\\
2	1.67419101387548\\
3	1.67419101387548\\
4	1.67419101387548\\
5	1.67419101387548\\
6	1.67419101387548\\
7	1.67419101387548\\
8	1.67419101387548\\
9	1.67419101387548\\
10	1.67419101387548\\
11	1.67419101387548\\
12	1.67419101387548\\
13	1.67419101387548\\
14	1.67419101387548\\
15	1.67419101387548\\
16	1.67419101387548\\
17	1.67419101387548\\
18	1.67419101387548\\
19	1.67419101387548\\
20	1.67419101387548\\
21	1.67419101387548\\
22	1.67419101387548\\
23	1.67419101387548\\
24	1.67419101387548\\
25	1.67419101387548\\
26	1.67419101387548\\
27	1.67419101387548\\
28	1.67419101387548\\
29	1.67419101387548\\
30	1.67419101387548\\
31	1.67419101387548\\
32	1.67419101387548\\
33	1.67419101387548\\
34	1.67419101387548\\
35	1.67419101387548\\
};
\addplot [color=black, line width=0.8pt, mark=*, mark size =1pt, mark options={solid, black}, forget plot]
  table[row sep=crcr]{%
1	1.51816337717656\\
2	1.57069402064057\\
3	1.59822577718413\\
4	1.61456083937775\\
5	1.62523074249726\\
6	1.63270567994323\\
7	1.63821954926654\\
8	1.64244899251058\\
9	1.64579347475647\\
10	1.64850318873694\\
11	1.65074252415981\\
12	1.6526238200227\\
13	1.654226388971\\
14	1.65560773234257\\
15	1.65681066364209\\
16	1.65786754577848\\
17	1.6588034487503\\
18	1.65963796851718\\
19	1.66038672797188\\
20	1.66106230884444\\
21	1.66167490812488\\
22	1.66223291677004\\
23	1.66274331915622\\
24	1.66321198947868\\
25	1.66364378974108\\
26	1.66404297491439\\
27	1.66441300778254\\
28	1.66475706820989\\
29	1.66507772436275\\
30	1.66537731120684\\
31	1.66565784212978\\
32	1.66592107114358\\
33	1.6661685346091\\
34	1.66640166278123\\
35	1.66662157904352\\
};
\node[above right, align=left]
at (axis cs:15,1.673) {\scriptsize $\sigma_2^2 = 10$};

 \addplot [color=red, line width=0.8pt, forget plot]
   table[row sep=crcr]{%
 1	1.858389650406\\
 2	1.858389650406\\
 3	1.858389650406\\
 4	1.858389650406\\
 5	1.858389650406\\
 6	1.858389650406\\
 7	1.858389650406\\
 8	1.858389650406\\
 9	1.858389650406\\
 10	1.858389650406\\
 11	1.858389650406\\
 12	1.858389650406\\
 13	1.858389650406\\
 14	1.858389650406\\
 15	1.858389650406\\
 16	1.858389650406\\
 17	1.858389650406\\
 18	1.858389650406\\
 19	1.858389650406\\
 20	1.858389650406\\
 21	1.858389650406\\
 22	1.858389650406\\
 23	1.858389650406\\
 24	1.858389650406\\
 25	1.858389650406\\
 26	1.858389650406\\
 27	1.858389650406\\
 28	1.858389650406\\
 29	1.858389650406\\
 30	1.858389650406\\
 31	1.858389650406\\
 32	1.858389650406\\
 33	1.858389650406\\
 34	1.858389650406\\
 35	1.858389650406\\
 };
 \addplot [color=black, line width=0.8pt, mark=*, mark size =1pt, mark options={solid, black}, forget plot]
   table[row sep=crcr]{%
 1	1.66401287907975\\
 2	1.73263559130195\\
 3	1.76735021599284\\
 4	1.78746984969687\\
 5	1.8004278924568\\
 6	1.80942567125765\\
 7	1.81602365508318\\
 8	1.82106340709723\\
 9	1.82503622516846\\
 10	1.82824732995557\\
 11	1.83089606581747\\
 12	1.83311781944442\\
 13	1.8350080138251\\
 14	1.83663557472007\\
 15	1.83805164022517\\
 16	1.83929479669557\\
 17	1.84039493076466\\
 18	1.84137531032198\\
 19	1.84225443469423\\
 20	1.84304729158603\\
 21	1.84376591343783\\
 22	1.84442029194741\\
 23	1.84501862817312\\
 24	1.84556783405371\\
 25	1.84607376572833\\
 26	1.8465412664421\\
 27	1.84697459275926\\
 28	1.84737737406472\\
 29	1.8477526724791\\
 30	1.84810326552925\\
 31	1.84843149701502\\
 32	1.84873944111796\\
 33	1.84902888699737\\
 34	1.84930149784899\\
 35	1.84955868120095\\
 };
 \node[above right, align=left]
 at (axis cs:15,1.858) {\scriptsize $\sigma_2^2 = 1000$};
\end{axis}

\end{tikzpicture}%
	\caption{Asymptotic behavior of $\Bar{\mathsf{R}}_n(1,\sigma_2^2)/\sqrt{n}$ versus $n$ for $\sigma_1^2 = 1$ and $\sigma_2^2 = 1.001,1.5,10,1000$.}
	\label{fig:asymRn}
\end{figure}
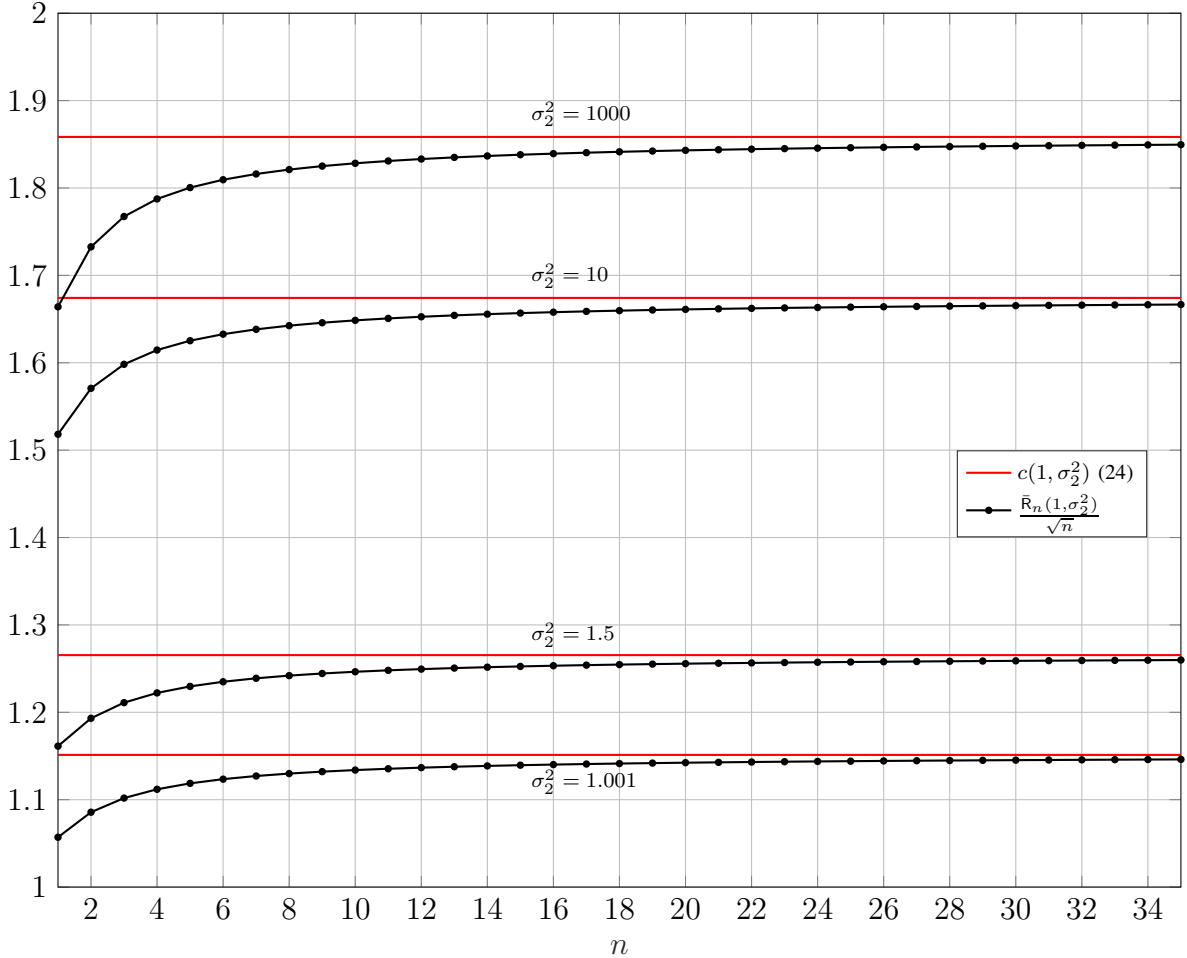%
In Fig.~\ref{fig:asymRn}, for $\sigma_1^2 = 1$ and $\sigma_2^2 = 1.001,1.5,10,1000$, we show the behavior of $\bar{\sfR}_n(1,\sigma_2^2)/\sqrt{n}$ and how its asymptotic converges to $c(1,\sigma_2^2)$.

\subsection{Scalar case $(n=1)$}

For the scalar case, we give an implicit and an explicit upper bound on the number of support points of the optimal input pmf $P_{X^{\star}}$.

\begin{theorem}\label{thm:Main_Results_Scalar}Let $Y_1^\star$ and $Y_2^\star$ be the secrecy-capacity-achieving output distributions at the legitimate and at the malicious receiver, respectively, and let
	\begin{align} \label{eq:functiongscalar}
	g(y)=\bbE\left[\log\frac{f_{Y_2^\star}(y+N)}{f_{Y_1^\star}(y)}\right],  \qquad  y\in \mathbb{R}, 
	\end{align} 
	with $N\sim {\cal N}(0,\sigma_2^2-\sigma_1^2)$.  For $\sfR>0$, an implicit upper bound on the number of support points of $P_{X^\star}$ is
	\begin{align}
	| \supp(P_{X^\star})| \le \rmN\left([-L,L], g(\cdot)+\kappa_1\right) <\infty \label{eq:Implicit_Upper_Bound_Scalar}
	\end{align}
	where 
	\begin{align}
	\kappa_1&=\log\left(\frac{\sigma_2}{\sigma_1}\right)-C_s,\\
	L&= \sfR \frac{\sigma_2+\sigma_1}{ \sigma_2-\sigma_1} +\sqrt{ \frac{ \frac{\sigma_2^2-\sigma_1^2}{\sigma_2^2}+2C_s}{ \frac{1}{\sigma_1^2}-\frac{1}{\sigma_2^2} } }. 
	\end{align}
	Moreover, an explicit upper bound on the number of support points of $P_{X^\star}$ is obtained by using  
	\begin{align}
	\rmN\left([-L,L], g(\cdot)+\kappa_1\right)  \le   \rho \frac{\sfR^2}{\sigma_1^2} + O( \log(\sfR) ), \label{eq:Upper_Bound_Explicit_Scalar}
	\end{align} 
	where $\rho= (2\rme+1)^2 \left( \frac{\sigma_2+\sigma_1}{ \sigma_2-\sigma_1} \right)^2+ \left(\frac{\sigma_2+\sigma_1}{ \sigma_2-\sigma_1}+1 \right)^2$.
\end{theorem} 
The upper bounds in Theorem~\ref{thm:Main_Results_Scalar} are generalizations of the upper bounds on the number of points presented in \cite{DytsoAmplitute2020} in the context of a point-to-point AWGN channel with an amplitude constraint.  Indeed, if we let $\sigma_2 \to \infty$, while keeping $\sigma_1$ and $\sfR$ fixed, then the wiretap channel reduces to the AWGN point-to-point channel. 

To find a lower bound on the number of mass points, a possible line of attack consists of the following steps: 
\begin{align}
C_s(\sigma_1^2, \sigma_2^2, \sfR, 1) &= I(X^\star;Y_1)- I(X^\star; Y_2)\\
& \le H(X^\star)- I(X^\star; Y_2)\\
& \le \log( | \supp(P_{X^\star})| ) - I(X^\star; Y_2), \label{eq:Step_1}
\end{align} 
where the above uses the nonnegativity of the entropy and the fact that entropy is maximized by a  uniform distribution.  Furthermore, by using a suboptimal uniform (continuous) distribution on $[-\sfR,\sfR]$ as an input and the entropy power inequality, the secrecy-capacity is lower-bounded by
\begin{equation}
C_s(\sigma_1^2, \sigma_2^2, \sfR, 1)  \ge   \frac{1}{2} \log \left( 1+ \frac{ \frac{2 \sfR^2}{ \pi \rme \sigma_1^2 } }{1+\frac{\sfR^2}{\sigma_2^2}} \right).  \label{eq:Step_2}
\end{equation} 
Combing the bounds in \eqref{eq:Step_1} and \eqref{eq:Step_2} we arrive at the following lower bound on the number of points:
\begin{equation}
| \supp(P_{X^\star})| \ge    \sqrt{1+ \frac{ \frac{2 \sfR^2}{ \pi \rme \sigma_1^2 } }{1+\frac{\sfR^2}{\sigma_2^2}}} \rme^{  I(X^\star; Y_2) } . 
\end{equation}
At this point one needs to determine the behavior of $I(X^\star; Y_2)$.  A trivial lower bound on  $ | \supp(P_{X^\star})| $ can be found by lower-bounding $ I(X^\star; Y_2)$ by zero. However, this lower bound  on $ | \supp(P_{X^\star})| $  does not grow with $\sfR$ while the upper bound does increase with $\sfR$. 
A possible way of establishing a lower bound that is increasing in $\sfR$ is by showing that $ I(X^\star; Y_2) \approx  \frac{1}{2} \log \left(1+\frac{\sfR^2}{\sigma_2^2} \right) $.  However, because not much is known about the structure of the optimal input distribution $P_{X^\star}$, it is not immediately evident how one can establish such an approximation or whether it is valid. 

\section{Secrecy-Capacity Expression in the Low Amplitude Regime} \label{sec:Cs_small_amp_regime} 
The result in Theorem~\ref{thm:Char_Small_Amplitude} can also be used to establish the secrecy-capacity for all $\sfR \le \bar{\sfR}_n(\sigma_1^2,\sigma_2^2)$ as is done next. 

\begin{theorem}\label{thm:Capacitiy_Small} If $\sfR \le \bar{\sfR}_n(\sigma_1^2,\sigma_2^2)$, then 
\begin{equation} \label{eq:Cs}
C_s(\sigma_1^2, \sigma_2^2, \sfR, n)=  \frac{1}{2} \int_{\sigma_1^2}^{\sigma_2^2} \frac{\sfR^2 -\sfR^2\bbE \left[      \mathsf{h}_{\frac{n}{2}}^2\left(  \frac{\|  \sfR+\sqrt{s}\bfZ\| \sfR}{s} \right) \right] }{s^2} \rmd s.
\end{equation}
\end{theorem} 
\begin{IEEEproof}
See Section~\ref{sec:thm:Capacitiy_Small}. 
\end{IEEEproof}

\subsection{Large $n$ Asymptotics }

Note that since $\bar{\sfR}_n(\sigma_1^2,\sigma_2^2)$ grows as $\sqrt{n}$, in view of Theorem~\ref{thm:large_n_beh}, then if we fix $\sfR$ and drive the number of antennas $n$ to infinity, the low amplitude regime becomes the only regime. The next theorem characterizes the secrecy-capacity in this `massive-MIMO' regime (i.e.,  where $\sfR$ is fixed and $n$ goes to infinity).
\begin{theorem}\label{thm:large_n_regime} Fix  $\sfR \ge 0$ and $\sigma_1^2 \le \sigma_2^2$, then
\begin{align}
&\lim_{n \to \infty} C_s(\sigma_1^2, \sigma_2^2, \sfR, n) =
  \sfR^2 \left( \frac{1}{2\sigma_1^2}- \frac{1}{2\sigma_2^2} \right).
\end{align} 
\end{theorem} 

\begin{IEEEproof}
To study the large $n$ behavior we will need to the following bounds on the function $ \mathsf{h}_{\nu}$ \cite{segura2011bounds,baricz2015bounds}: for $\nu > \frac{1}{2}$  
\begin{align}
 \mathsf{h}_{\nu}(x)=   \frac{x}{ \frac{2\nu-1}{2}+\sqrt{ \frac{(2\nu-1)^2}{4} +x^2}} \cdot g_\nu(x), \label{eq:expression_for_h}
\end{align}
where 
\begin{align}
1 \ge g_\nu(x) \ge    \frac{ \frac{2\nu-1}{2}+\sqrt{ \frac{(2\nu-1)^2}{4} +x^2}}{ \nu+\sqrt{ \nu^2 +x^2} }.
\end{align} 
Moreover, let
\begin{equation}
U_n = \|  \sfR+\sqrt{s}\bfZ\|
\end{equation} 
with $\bfZ \sim \mathcal{N}(\mathbf{0}_n,\sigma^2 \bfI_n)$.
Consequently,
\begin{align}
&\lim_{n \to \infty} \bbE \left[      \mathsf{h}_{\frac{n}{2}}^2\left(  \frac{\|  \sfR+\sqrt{s}\bfZ\| \sfR}{s} \right) \right]\\
&=\bbE \left[  \lim_{n \to \infty}      \mathsf{h}_{\frac{n}{2}}^2\left(  \frac{\|  \sfR+\sqrt{s}\bfZ\| \sfR}{s} \right) \right] \label{eq:Applying_DCT1}\\
&=  \bbE \left[   \lim_{n \to \infty}   \frac{U_n^2 \frac{\sfR^2}{s^2}}{ \left( \frac{n-1}{2}+\sqrt{ \frac{(n-1)^2}{4} +U_n^2 \frac{\sfR^2}{s^2} } \right)^2} \cdot g_{ \frac{n}{2}}^2\left(U_n \frac{\sfR}{s} \right) \right] \label{eq:applying_approx_h}\\
&= \bbE \left[   \lim_{n \to \infty}   \frac{ \frac{1}{n}U_n^2 \frac{\sfR^2}{s^2}}{ n \cdot \left( \frac{1}{2}+\sqrt{ \frac{1}{4} +\left( \frac{1}{n}U_n \frac{\sfR}{s} \right)^2} \right)^2} \cdot g_{ \frac{n}{2}}^2 \left(U_n \frac{\sfR}{s} \right) \right]\\
&= 0, \label{eq:using_SLLN}
\end{align} 
where \eqref{eq:Applying_DCT1} follows from the dominated convergence theorem since $|h_\nu| \le 1$;  \eqref{eq:applying_approx_h} follows from using   \eqref{eq:expression_for_h}; \eqref{eq:using_SLLN} follows from using the strong law of large numbers to note that  
\begin{equation}
\lim_{n \to \infty} \frac{1}{n} U_n^2=\lim_{n \to \infty} \frac{\|  \sfR+\sqrt{s}\bfZ\|^2}{n}=s.
\end{equation} 
Now, combining the capacity expression in  \eqref{eq:Cs} and \eqref{eq:using_SLLN} we have that 
\begin{align}
\lim_{n \to \infty} C_s(\sigma_1^2, \sigma_2^2, \sfR, n) =\frac{1}{2} \int_{\sigma_1^2}^{\sigma_2^2} \frac{\sfR^2}{s^2} \rmd s= \sfR^2 \left( \frac{1}{2\sigma_1^2}- \frac{1}{2\sigma_2^2} \right).
\end{align}

\end{IEEEproof}

\begin{rem}
The result in Theorem~\ref{thm:large_n_regime}, is reminiscent of the capacity in the wideband regime \cite[Ch.~9]{Cover:InfoTheory} where the capacity increases linearly in the signal-to-noise ratio.  Similarly, Theorem~\ref{thm:large_n_regime} shows that in the large antenna regime the secrecy-capacity grows linearly as the difference of the single-to-noise ratio at the legitimate user and at the eavesdropper. 
\end{rem}

In Theorem~\ref{thm:large_n_regime},   $\sfR$ was held fixed. It is also interesting to study the case when $\sfR$ is a function of $n$. Specifically, it is interesting to study the case when $\sfR= c \sqrt{n}$ for some coefficient $c$.

\begin{theorem}\label{thm:large_n_coupled} Suppose that $c \le c(\sigma_1^2,\sigma_2^2)$. Then, 
\begin{equation} \label{eq:Capacity_for_large_n}
 \lim_{n \to \infty} \frac{C_s(\sigma_1^2, \sigma_2^2, c \sqrt{n},n )}{n} =  \frac{1}{2} \log \left( \frac{1+c^2/\sigma_1^2}{ 1+c^2/\sigma_2^2} \right). 
\end{equation}
\end{theorem}

\begin{IEEEproof} Let $\sfR_n =c \sqrt{n}$
\begin{align}
&\lim_{n \to \infty} \frac{C_s(\sigma_1^2, \sigma_2^2, \sfR_n,n)}{n} \notag\\
&=  \frac{c^2}{2} \int_{\sigma_1^2}^{\sigma_2^2} \frac{1 - \lim_{n \to \infty} \bbE \left[      \mathsf{h}_{\frac{n}{2}}^2\left(  \frac{\|  \sfR_n+\sqrt{s}\bfZ\| \sfR_n}{s} \right) \right] }{s^2} \rmd s\\
&=  \frac{c^2}{2} \int_{\sigma_1^2}^{\sigma_2^2} \frac{1 - \frac{ c^2 (c^2+ s)}{ \left( \frac{s}{2}+\sqrt{ \frac{s^2}{4} +c^2( c^2+ s)  } \right)^2} }{s^2} \rmd s \label{eq:lim_hv} \\
&= \frac{1}{2} \log \left( \frac{\sigma_2^2 (c^2+\sigma_1^2)}{ \sigma_1^2(c^2+\sigma_2^2)} \right),
\end{align}
where \eqref{eq:lim_hv} follows from the limit established in \eqref{eq:using_Law_large_numbers_second}. This concludes the proof. 
\end{IEEEproof} 
The result in \eqref{eq:Capacity_for_large_n} can be recast as follows.  Consider the secrecy-capacity of vector Gaussian wiretap channel subject to the average-power constraint (i.e., $\mathbb{E}[\| \bfX\|^2] \le \mathsf{P}$):
\begin{equation}
C_G(\sigma_1^2,\sigma_2^2, \mathsf{P},n) = \frac{n}{2} \log \frac{1+\mathsf{P}/\sigma_1^2}{1+\mathsf{P}/\sigma_2^2}.
\end{equation} 
Thus, the result in \eqref{eq:Capacity_for_large_n} can be restated as
\begin{equation}
\lim_{n \to \infty} \frac{C_s(\sigma_1^2, \sigma_2^2, c\sqrt{n},n)}{C_G(\sigma_1^2,\sigma_2^2, c^2,n) }=1.
\end{equation}
In other words,  for the regime considered in Theorem~\ref{thm:large_n_coupled}, for large enough $n$ the secrecy-capacity under the amplitude constraint $\sfR_n =c \sqrt{n}$ behaves as the secrecy-capacity under the average power constraint $c^2$.

\section{Beyond the Low Amplitude Regime} \label{sec:beyond_small_amp_regime}
To evaluate the secrecy-capacity and find the optimal distribution $P_{\bfX^\star}$ beyond $\bar{\sfR}_n$ we rely on numerical estimations. We remark that, as pointed out in \cite{DytsoITWwiretap2018}, the secrecy-capacity-achieving distribution is isotropic and consists of finitely many co-centric shells. Keeping this in mind, we can find the optimal input distribution $P_{\bfX^\star}$ by just optimizing over $P_{\|\bfX \|}$ with $\|\bfX \|\le \sfR$.

\subsection{Numerical Algorithm}
	
	Let us denote by $\hat{C}_s(\sigma_1^2,\sigma_2^2,\sfR,n)$ the numerical estimate of the secrecy-capacity and by $\hat{P}_{\|\bfX^\star\|}$ the estimate of the optimal pmf on the input norm. To numerically evaluate $\hat{C}_s(\sigma_1^2,\sigma_2^2,\sfR,n)$ and $\hat{P}_{\|\bfX^\star\|}$ we rely on an algorithmic procedure similar to the one described in~\cite{barletta2021numerical}, which in turn takes inspiration from the deterministic annealing algorithm sketched in~\cite{rose1994mapping}. The numerical procedure is given in Algorithm~\ref{alg:CapacityEst}. The input parameters of the main function are the noise variances $\sigma_1^2$ and $\sigma_2^2$, the radius $\sfR$, the vectors $\bfrho$ and $\bfp$ being, respectively, the mass points positions and probabilities of a tentative input pmf, the number of iterations in the while loop $N_c$, and finally a tolerance $\varepsilon$ to set the precision of the secrecy-capacity estimate.
	\begin{figure}[!t]
		\removelatexerror
		\begin{algorithm}[H]
			\caption{Secrecy-capacity and optimal input pmf estimation}
			\label{alg:CapacityEst}
			\begin{algorithmic}[1]
				\Procedure{Main}{$\left( \sigma_1^2, \sigma_2^2,\sfR, \bfrho , \bfp ,N_c, \varepsilon \right)$} \vspace*{0.35ex}
				\Repeat
				\State $k \gets 0$
				\While{$k < N_c$}
				\State $k \gets k+1$
				\State $\bfrho \gets \textsc{Gradient-Ascent}(\bfrho , \bfp)$
				\State $\bfp \gets \textsc{Blahut-Arimoto}(\bfrho , \bfp)$
				\EndWhile
				\State valid $\gets \textsc{KKT-Validation}(\bfrho , \bfp,\varepsilon) $
				\If{valid $=$ False}
				\State $(\bfrho , \bfp) \gets \textsc{Add-Point}(\bfrho , \bfp )$
				\EndIf
				\Until valid $=$ True \vspace*{0.7ex}
				\State $\hat{P}_{\|\bfX^\star \|} \gets (\bfrho , \bfp )$ \vspace*{0.7ex}
				\State $\hat{C}_s\left(\sigma_1^2, \sigma_2^2,\sfR,n \right) \gets \Xi\left(\sfR ; \hat{P}_{\|\bfX^\star \|}\right) $ \vspace*{0.7ex}
				\State \Return $\hat{P}_{\|\bfX^\star \|},\hat{C}_s\left(\sigma_1^2, \sigma_2^2,\sfR,n \right)$ \vspace*{0.7ex}
				\EndProcedure
			\end{algorithmic}
		\end{algorithm}
		\vspace*{-6ex}
	\end{figure}
	At its core the numerical procedure iteratively refines its estimate of $P_{\|\bfX^\star\|}$ by running a gradient ascent algorithm to update the vector $\bfrho$ and a variant of the Blahut-Arimoto algorithm~\cite{blahut1972computation} to update~$\bfp$.
	
	The \textsc{Gradient-Ascent} procedure uses the secrecy-information $I(\bfX; \bfY_1) - I(\bfX; \bfY_2)$ as objective function and stops either when $\bfrho$ has reached convergence or at a given maximum number of iterations. We remark that to ensure the convergence to a local maximum, we use the gradient-ascent algorithm in a backtracking line search version~\cite{boyd2004convex}. The backtracking line search version guarantees us that each new update of $\bfrho$ provides a nondecreasing associated secrecy-information, compared to the previous update of $\bfrho$. 
	
	Similarly to \textsc{Gradient-Ascent}, the \textsc{Blahut-Arimoto} procedure stops either when the values of $\bfp$ have reached a stable convergence or after a set number of updates.
	
	Since the joint optimization of $\bfrho$ and $\bfp$ is not numerically feasible, we need to reiterate both the \textsc{Blahut-Arimoto} and the \textsc{Gradient-Ascent} procedures a given number of times, namely $N_c$. The parameter $N_c$ is chosen empirically in such a way that $\bfrho$ and $\bfp$ become fairly stable and therefore we can expect to have reached joint convergence for both of them.
	
	 Then, the \textsc{KKT-Validation} procedure ensures that the values of $\bfrho$ and $\bfp$ are indeed close to the optimal ones. Let us denote by $\hat{P}_{\| \bfX \|}$ the tentative pmf of mass points $\bfrho$ and corresponding probabilities $\bfp$. We check the optimality of $\hat{P}_{\| \bfX \|}$ by verifying whether the KKT conditions in Lemma~\ref{lem:KKT} are satisfied. Since the algorithm has to verify the KKT conditions numerically, i.e., with finite precision, we find more convenient to check the negated version of~\eqref{eq:NewOptimalityEquations}, where a tolerance parameter $\varepsilon$ is introduced which trades off accuracy with computational burden. Specifically, $\hat{P}_{\| \bfX \|}$ is not an optimal input pmf if any of the following conditions is satisfied:
	\begin{subequations}
	\begin{align}
		 &| \Xi(t;\hat{P}_{\| \bfX \|}) - \Xi(\sfR;\hat{P}_{\| \bfX \|}) | > \varepsilon, \quad \text{for some } t \in \supp(\hat{P}_{\| \bfX \|}) \label{eq:epsKKT1} \\
		   &\Xi(\sfR;\hat{P}_{\| \bfX \|})+\varepsilon < \Xi(t;\hat{P}_{\| \bfX \|}) , \quad \text{for some }t \in [0,\sfR], \label{eq:epsKKT2}
	\end{align}
	\label{eq:epsKKT}
\end{subequations}
	where $\Xi(t;\hat{P}_{\| \bfX \|})$ is the secrecy-density, with respect to the input norm, defined in~\eqref{eq:secrecy-density_norm}.
	 Note that in~\eqref{eq:epsKKT} in place of the secrecy-capacity $C_s(\sigma_1^2, \sigma_2^2,\sfR,n)$, which is unknown, we used the value of $\Xi(\sfR;\hat{P}_{\| \bfX \|})$, thanks to the fact that $\sfR \in \supp(P_{\|\bfX^\star\|})$ for any~$(\sigma_1,\sigma_2, \sfR, n)$. Condition~\eqref{eq:epsKKT1} is derived by negating~\eqref{eq:EqualityCOndition}: there exists a $t \in \supp(\hat{P}_{\| \bfX \|})$ such that $\Xi(t;\hat{P}_{\| \bfX \|})$ is $\varepsilon$-away from the estimated secrecy-capacity $\Xi(\sfR;\hat{P}_{\| \bfX \|})$. Condition~\eqref{eq:epsKKT2} is the negated version of~\eqref{eq:not_larger_than_C}: there exists a $t \in [0,\sfR]$ such that $\Xi(t;\hat{P}_{\| \bfX \|})$ is at least $\varepsilon$-larger than the estimated secrecy-capacity $\Xi(\sfR;\hat{P}_{\| \bfX \|})$.  With some abuse of notation, we refer to~\eqref{eq:epsKKT} as to the $\varepsilon$-KKT conditions.
	 If the tentative pmf $\hat{P}_{\| \bfX \|}$ does not pass the check of the $\varepsilon$-KKT conditions, then the algorithm checks whether a new point has to be added to the pmf.
	
	The \textsc{Add-Point} procedure evaluates the position of the new mass point
	\begin{align}
		\rho_{\text{new}} = \arg \max_{t \in [0,\sfR]} \Xi(t;\hat{P}_{\| \bfX \|}).
	\end{align}
	The point $\rho_{\text{new}}$ is appended to the vector $\bfrho$ and the probabilities $\bfp$ are set to be equiprobable.
	
	The whole procedure is repeated until \textsc{KKT-Validation} gives a positive outcome and at that point the algorithm returns $\hat{P}_{\| \bfX \|}$ as the optimal pmf estimate and $\hat{C}_s(\sigma_1^2, \sigma_2^2,\sfR ,n)$ as the secrecy-capacity estimate.

	\begin{rem}
		In this work we focus on the secrecy-capacity and on the secrecy-capacity-achieving input distribution. However, it is possible to study other points of the rate-equivocation region of the degraded wiretap Gaussian channel by suitably changing the KKT conditions as reported in~\cite[Eq.~(33) and (34)]{ozel2015gaussian}. With the due modifications, the proposed optimization algorithm can find the optimal input distribution for any point of the rate-equivocation region.
	\end{rem}

	\subsection{Numerical Results}
\begin{figure}[t]
	\centering
	\input{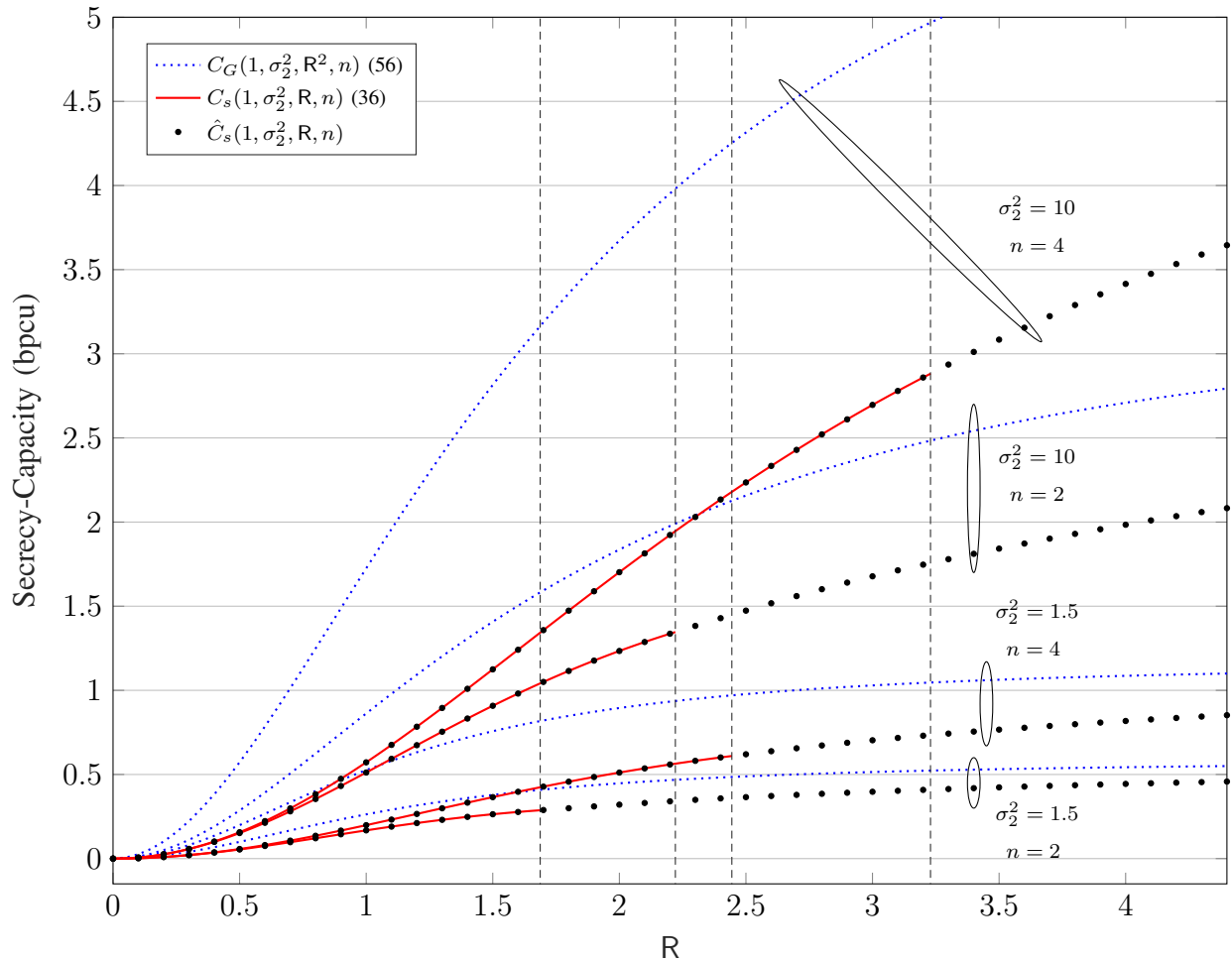}
	\caption{Secrecy-capacity in bit per channel use (bpcu) versus $\sfR$, for $\sigma_2^2 = 1.5,10$ and $n=2,4$.}\label{fig:Cs_vs_Cnum}
\end{figure}

	In Fig.~\ref{fig:Cs_vs_Cnum}, we show with black dots the numerical estimate $\hat{C}_s(\sigma_1^2,\sigma_2^2,\sfR,n)$ versus $\sfR$, evaluated via Algorithm~\ref{alg:CapacityEst}, for $\sigma_1^2 = 1$, $\sigma_2^2 = 1.5, 10$, $n=2,4$, and tolerance $\varepsilon = 10^{-6}$. For the same values of $\sigma_1^2$, $\sigma_2^2$, and $n$ we also show, with the red lines, the analytical low amplitude regime secrecy-capacity $C_s(\sigma_1^2,\sigma_2^2,\sfR,n)$ versus $\sfR$ from Theorem~\ref{thm:Capacitiy_Small}. Also, we show with blue dotted lines the secrecy-capacity under the average-power constraint $\bbE \left[ \| \bfX \|^2 \right] \leq \sfR^2$:
	\begin{align} \label{eq:C_G}
		 C_G(\sigma_1^2,\sigma_2^2,\sfR^2,n) &= \frac{n}{2} \log \frac{1+\sfR^2/\sigma_1^2}{1+\sfR^2/\sigma_2^2}\ge C_s(\sigma_1^2,\sigma_2^2,\sfR,n),
	\end{align}
	where the inequality follows by noting that the average-power constraint $\bbE \left[ \| \bfX \|^2 \right] \leq \sfR^2$ is weaker than the amplitude constraint $\| \bfX \| \leq \sfR$. Finally, the dashed vertical lines show $\bar{\sfR}_n$,  i.e. the upper limit of the low amplitude regime, for the considered values of $\sigma_1^2$, $\sigma_2^2$, and $n$.
	
	In Fig.~\ref{fig:PMFevo_n2},  we consider discrete values for $\sfR$ and for each value of $\sfR$  we plot the corresponding estimated pmf $\hat{P}_{\|\bfX^\star\|}$, evaluated via Algorithm~\ref{alg:CapacityEst}, for $\sigma_1^2 = 1$, $\sigma_2^2 = 1.5$, $n=2,8$, and tolerance $\varepsilon = 10^{-6}$. The figure shows, at each $\sfR$, the normalized amplitude  of support points in the estimated pmf, while the size of the circles qualitatively shows the probability  associated with each support point. Similarly, Fig.~\ref{fig:PMFevo_n2v10} shows the evolution of the pmf estimate for $\sigma_1^2 = 1$, $\sigma_2^2 = 10$, $n=2,8$, and $\varepsilon = 10^{-6}$. It is interesting to notice how in both Fig.~\ref{fig:PMFevo_n2} and Fig.~\ref{fig:PMFevo_n2v10} when a new mass point is added to the pmf, it appears in zero.
	
\begin{figure*}[t]
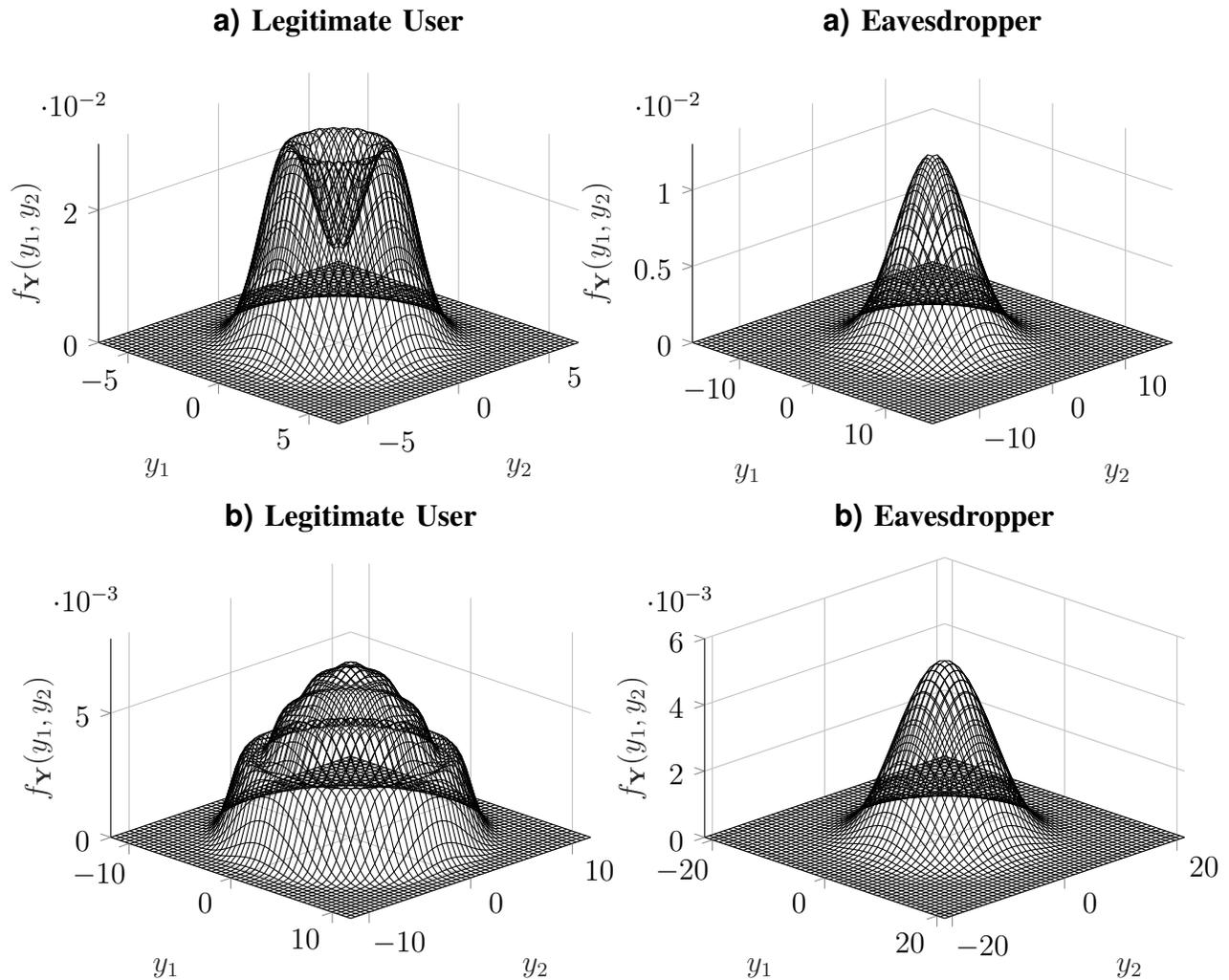

	\centering
	\input{Figures/pdfY_R=2.25_var2=10} 
	\input{Figures/pdfY_R=7.5_var2=10}  
	\caption{Output pdf of the legitimate user and of the eavesdropper for $\sigma_1^2 = 1$, $\sigma_2^2 = 10$, $n=2$, \textbf{\textsf{a)}} $\sfR = 2.25$, and \textbf{\textsf{b)}}  $\sfR = 7.5$.  An animation showing the evolution of the output pdf, as $\sfR$ varies, can be found in~\cite{GithubData}. }	
	\label{fig:pdfY}
\end{figure*}

	Finally, Fig.~\ref{fig:pdfY} shows the output distributions of the legitimate user and of the eavesdropper in the case of $\sigma_1^2 = 1$, $\sigma_2^2 = 10$, $n = 2$, and for two values of $\sfR$. At the top of the figure, the distributions are shown for $\sfR = 2.25$, which is a value close to $\bar{\sfR}_2(1,10)$. At the bottom of the figure, the distributions are shown for $\sfR = 7.5$. For both values of $\sfR$, the legitimate user sees an output distribution where the co-centric rings of the input distribution are easily distinguishable. On the other hand, as expected, the output distribution seen by the eavesdropper is close to a Gaussian.

\begin{figure}[t]
	\centering
	\input{Figures/PMFevolution_n=2_var2=1.5}
	\caption{Evolution of the numerically estimated $\hat{P}_{\|\bfX^\star\|}$ versus $\sfR$ for $\sigma_1^2 = 1$, $\sigma_2^2 = 1.5$, \textbf{\textsf{a)}} $n=2$, and \textbf{\textsf{b)}} $n=8$.}
	\label{fig:PMFevo_n2}
\end{figure}

\begin{figure}[t]
	\centering
	\input{Figures/PMFevolution_n=2_var2=10}
	\caption{Evolution of the numerically estimated $\hat{P}_{\|\bfX^\star\|}$ versus $\sfR$ for $\sigma_1^2 = 1$, $\sigma_2^2 = 10$, \textbf{\textsf{a)}} $n=2$, and \textbf{\textsf{b)}} $n=8$.} 
	\label{fig:PMFevo_n2v10}
\end{figure}

\section{Proof of Theorem~\ref{thm:Char_Small_Amplitude}} \label{sec:thm:Char_Small_Amplitude}

\subsection{KKT Conditions}
\begin{lemma}\label{lem:KKT} $P_{\bfX^\star}$ maximizes \eqref{eq:Secracy_CAP} if and only if 
\begin{subequations}
\begin{align}
\Xi(\bfx;P_{\bfX^\star}) &= C_s(\sigma_1^2, \sigma_2^2,\sfR,n), \qquad \bfx \in \supp(P_{\bfX^\star}),  \label{eq:EqualityCOndition}\\
\Xi(\bfx;P_{\bfX^\star}) &\le C_s(\sigma_1^2, \sigma_2^2, \sfR,n), \qquad \bfx \in  \cB_0(\sfR) ,\label{eq:not_larger_than_C}
\end{align}
	\label{eq:NewOptimalityEquations}
\end{subequations}
where  for $\bfx \in \mathbb{R}^n$
\begin{align}
\Xi(\bfx;P_{\bfX^\star})&=\sfD(f_{\bfY_1|\bfX}(\cdot| \bfx) \|f_{\bfY_1^\star})- \sfD(f_{\bfY_2|\bfX}(\cdot|\bfx) \|f_{\bfY_2^\star}) \label{eq:secrecy-density} \\
&=\bbE \left[ g(\bfY_1) | \bfX=\bfx \right] , \label{eq:Writing_KKT_as_statistics}
\end{align}
and where 
\begin{align} \label{eq:functiong}
g(\bfy)=\bbE\left[\log\frac{f_{\bfY_2^\star}(\bfy+\bfN)}{f_{\bfY_1^\star}(\bfy)}\right]+ n \log\left(\frac{\sigma_2}{\sigma_1}\right),  \,  \bfy\in \mathbb{R}^n, 
\end{align} 
with $\bfN\sim {\cal N}(\mathbf{0}_n, (\sigma_2^2-\sigma_1^2) \bfI_n )$. 
\end{lemma} 
\begin{proof}
	This is a vector extension of Lemma~\ref{lem:KKT_Scalar}, which is presented in Section~\ref{Sec:KKT_Scalar}.
\end{proof}

\subsection{A New Necessary and Sufficient Condition}

\begin{theorem}\label{thm:equivalent_condition} $P_{\bfX_{\sfR}} $ is optimal if and only if for all $\| \bfx \|=\sfR$
\begin{equation}\label{eq:ineq_density}
\Xi({\bf 0};P_{\bfX_{\sfR}}) \le \Xi(\bfx;P_{\bfX_{\sfR}}). 
\end{equation} 
Moreover, if 
\begin{equation}
\sfR < \sigma_1^2 \sqrt{n \left(\frac{1}{\sigma_1^2}-\frac{1}{\sigma_2^2}\right)}
\end{equation}
then $P_{\bfX_{\sfR}} $ is optimal.
\end{theorem} 

\begin{proof}
	The secrecy-density $\Xi(\cdot; P_{\bfX_{\sfR}})$ is a function only of $\| \bfx\|$, thanks to the rotational symmetry of the  additive noise distribution and of $P_{\bfX_{\sfR}}$. In view of this, a way to prove condition~\eqref{eq:ineq_density} is to show that the maximum of $\|\bfx\|\mapsto\Xi(\|\bfx\|;P_{\bfX_{\sfR}})$ occurs at either $\|\bfx\|=0$ or $\|\bfx\|=\sfR$. Next, we show that the derivative of $\Xi(\|\bfx\|;P_{\bfX_{\sfR}})$ makes at most one sign change, from negative to positive. This fact will prove the claim.
	
	From Lemma \ref{Lemma:derivative_Xi} in the Appendix, the derivative of $\Xi$ is\footnote{A related calculation was erroneously performed in \cite{dytsoMI_est_2019}. However, this error does not change the results of \cite{dytsoMI_est_2019} as only the sign of the derivative is important and not the value itself.} 
	\begin{align} \label{eq:derivative_Xi}
	\Xi'(\|\bfx\|;P_{\bfX_{\sfR}}) =\|\bfx \|\:\bbE\left[\widetilde{M}_2(\sigma_1 Q_{n+2})-M_1(\sigma_1 Q_{n+2}) \right]
	\end{align}
	where $Q_{n+2}^2$ is a noncentral chi-square random variable with $n+2$ degrees of freedom and noncentrality parameter $\frac{\|\bfx\|^2}{\sigma_1^2}$ and
	\begin{align}
	M_i(y) &= \frac{1}{\sigma_i^2}\left(\frac{\sfR}{y}\sfh_{  \frac{n}{2} }\left(\frac{\sfR}{\sigma_i^2}y\right)-1\right), \qquad i\in \{1,2\} \\
	\widetilde{M}_2(y) &= \bbE\left[M_2(\|y+\bfW \|)\right],
	\end{align}
	where $\bfW \sim {\cal N}(\mathbf{0}_{n+2},(\sigma_2^2-\sigma_1^2)\bfI_{n+2})$.

Note that $\Xi'(0;P_{\bfX_{\sfR}})=0$, and that $\Xi'(\|\bfx\|;P_{\bfX_{\sfR}})>0$ for sufficiently large $\|\bfx\|$; in fact, we have
\begin{align}
&\Xi'(\|\bfx\|;P_{\bfX_{\sfR}}) > \|\bfx\|\left(\frac{1}{\sigma_1^2}-\frac{1}{\sigma_2^2}\right) -\frac{\|\bfx\|}{\sigma_1^2}\bbE\left[\frac{\sfR}{\sigma_1 Q_{n+2}}  \right] \label{eq:applyboundsonh} \\
&=\|\bfx\|\left(\frac{1}{\sigma_1^2}-\frac{1}{\sigma_2^2}\right) -\frac{\|\bfx\|}{\sigma_1^2}\bbE\left[\frac{\sfR}{\|\bfx\|} \sfh_{\frac{n}{2}}\left(\frac{\|\bfx\|}{\sigma_1}Q_n\right) \right] \label{eq:ndegreesoffreedom} \\
&\ge\|\bfx\|\left(\frac{1}{\sigma_1^2}-\frac{1}{\sigma_2^2}\right) -\frac{\sfR}{\sigma_1^2}, \label{eq:applyboundsonh1}
\end{align} 
where \eqref{eq:applyboundsonh} follows from $0\le \sfh_{\frac{n}{2}}(x) \le 1$ for $x\ge 0$; \eqref{eq:ndegreesoffreedom} follows from a change of measure in the expectation; and finally \eqref{eq:applyboundsonh1} holds by $\sfh_{\frac{n}{2}}(x) \le 1$.

 To conclude, we need to prove that $\Xi'(\|{\bfx} \|;P_{\bfX_{\sfR}})$ changes sign at most once. To that end, we will need Karlin's oscillation theorem presented in Sec.~\ref{sec:oscillation}. 
 By using~\eqref{eq:derivative_Xi}, the fact that the pdf of a chi-square is positive defined kernel~\cite{karlin1957polya}, and Theorem~\ref{thm:OscillationThoerem}, the number of sign changes of $\Xi'(\|{\bfx} \|;P_{\bfX_{\sfR}})$ is upper-bounded by the number of sign changes of 
\begin{equation}
\widetilde{M}_2(y)-M_1(y)=G_{\sigma_1,\sigma_2,\sfR,n}(y), 
\end{equation}
for $y>0$ where $G_{\sigma_1,\sigma_2,\sfR,n}(y)$ was defined and discussed in Section~\ref{sec:Assumptions} and it was assumed that it has at most one sign change for $y >0$. For example, a sufficient condition is given by 
\begin{equation}
\sfR < \sigma_1^2 \sqrt{n \left(\frac{1}{\sigma_1^2}-\frac{1}{\sigma_2^2}\right)}
\end{equation}
This concludes the proof.

\end{proof}

\subsection{Estimation Theoretic Representation}

To complete the proof we seek to re-write the condition in Theorem~\ref{thm:equivalent_condition} in the estimation theoretic form. To that end, we need the following representation of the relative entropy~\cite{verdu2010mismatched}: 
\begin{equation}
\sfD(P_{\bfX_1+\sqrt{t}\bfZ} \| P_{\bfX_2+\sqrt{t}\bfZ})
= \frac{1}{2} \int_t^\infty  \frac{g(s)}{s^2} \rmd s , \label{eq:mistmatchedMMSE_formulat}
\end{equation} 
where 
\begin{align}
g(s)&= \bbE \left[ \| \bfX_1 -\phi_2(\bfX_1+\sqrt{s} \bfZ)    \|^2 \right]  - \bbE \left[ \| \bfX_1 -\phi_1(\bfX_1+\sqrt{s} \bfZ)    \|^2 \right],
\end{align} 
and where
\begin{align}
\phi_i(\bfy)=\bbE[\bfX_i|\bfX_i+\sqrt{s} \bfZ=\bfy], \,  i\in \{1,2\}. 
\end{align} 

Another fact that will be important for our expression is 
\begin{align}
\bbE \left[\bfX_{\sfR} \mid \bfX_{\sfR}+\sqrt{s} \bfZ=\bfy \right]=   \frac{\sfR \bfy}{\|\bfy\|}    \mathsf{h}_{\frac{n}{2}}\left(  \frac{\| \bfy\| \sfR}{s} \right),  \label{eq:ConditionalExpectation}
\end{align} 
see, for example \cite{dytsoMI_est_2019}, for the proof. 

Next, using \eqref{eq:mistmatchedMMSE_formulat} and \eqref{eq:ConditionalExpectation} note that for any $\| \bfx\|=\sfR$ we have that for $i \in \{1,2\}$
\begin{align}
& \sfD(P_{ {\bfx}+\sqrt{\sigma^2_i}\bfZ} \| P_{\bfX_{\sfR}+\sqrt{\sigma^2_i}\bfZ})\\
 &=\frac{1}{2} \int_{\sigma_i^2}^\infty  \frac{ \bbE \left[  \left\| \bfx -  \frac{\sfR ( {\bfx}+\sqrt{s}\bfZ)}{\| {\bfx}+\sqrt{s}\bfZ\|}    \mathsf{h}_{\frac{n}{2}}\left(  \frac{\|  {\bfx}+\sqrt{s}\bfZ\| \sfR}{s} \right) \right\|^2\right] }{s^2} \rmd s\\
  &=\frac{1}{2} \int_{\sigma_i^2}^\infty  \frac{\sfR^2 -\sfR^2\bbE \left[      \mathsf{h}_{\frac{n}{2}}^2\left(  \frac{\|  {\bfx}+\sqrt{s}\bfZ\| \sfR}{s} \right)\right] }{s^2} \rmd s, \label{eq:KL_at_x}
\end{align}
and
\begin{align}
\sfD(P_{ {\bf 0}+\sqrt{\sigma^2_i}\bfZ} \| P_{\bfX_{\sfR}+\sqrt{\sigma^2_i}\bfZ})=\frac{1}{2} \int_{\sigma_i^2}^\infty  \frac{ \sfR^2 \bbE \left[ \sfh_{  \frac{n}{2} }^2 \left( \frac{ \sfR \|\bfZ\|}{s} \right)\right] }{s^2} \rmd s. \label{eq:KL_at_0}
\end{align}

Now, note that by using definition of $\Xi(\bfx; P_{\bfX_{\sfR}})$ in~\eqref{eq:Writing_KKT_as_statistics}, and \eqref{eq:KL_at_x} and~\eqref{eq:KL_at_0} we have that for $\| \bfx\|=\sfR$
\begin{align}
&\Xi(\bfx; P_{\bfX_{\sfR}})\notag\\
&=\sfD(P_{ {\bfx}+\sqrt{\sigma^2_1}\bfZ} \| P_{\bfX_{\sfR}+\sqrt{\sigma^2_1}\bfZ})- \sfD(P_{ {\bf x}+\sqrt{\sigma^2_2}\bfZ} \| P_{\bfX_{\sfR}+\sqrt{\sigma^2_2}\bfZ})\\
&=\frac{1}{2} \int_{\sigma_1^2}^{\sigma_2^2} \frac{\sfR^2 -\sfR^2\bbE \left[      \mathsf{h}_{\frac{n}{2}}^2\left(  \frac{\|  {\bfx}+\sqrt{s}\bfZ\| \sfR}{s} \right) \right] }{s^2} \rmd s, \label{eq:diff_of_KL_s}
\end{align} 
and 
\begin{align}
&\Xi({\bf 0} ;P_{\bfX_{\sfR}})\notag\\
&=\sfD(P_{ {\bf 0}+\sqrt{\sigma^2_1}\bfZ} \| P_{\bfX_{\sfR}+\sqrt{\sigma^2_1}\bfZ})-  \sfD(P_{ {\bf 0}+\sqrt{\sigma^2_2}\bfZ} \| P_{\bfX_{\sfR}+\sqrt{\sigma^2_2}\bfZ})\\
&=\frac{1}{2} \int_{\sigma_1^2}^{\sigma_2^2} \frac{\sfR^2\bbE \left[      \mathsf{h}_{\frac{n}{2}}^2\left(  \frac{\|  \sqrt{s}\bfZ\| \sfR}{s} \right) \right] }{s^2} \rmd s
\end{align}

Consequently, the necessary and sufficient condition in Theorem~\ref{thm:equivalent_condition} can be equivalently written as
\begin{align}
& \int_{\sigma_1^2}^{\sigma_2^2} \frac{\bbE \left[      \mathsf{h}_{\frac{n}{2}}^2\left(  \frac{\|  \sqrt{s}\bfZ\| \sfR}{s} \right) +     \mathsf{h}_{\frac{n}{2}}^2\left(  \frac{\|  {\bfx}+\sqrt{s}\bfZ\| \sfR}{s} \right) \right]-1}{s^2} \rmd s \le 0.\label{eq:Final_inequality_thm1} 
\end{align} 

Now $\bar{\sfR}_n(\sigma_1^2,\sigma_2^2)$ will be the largest $\sfR$ that satisfies \eqref{eq:Final_inequality_thm1}, which concludes the proof of Theorem~\ref{thm:Char_Small_Amplitude}.

\section{Proof of Theorem~\ref{thm:large_n_beh}} 
\label{sec:large_n_beh}
The objective of the proof is to understand how the condition in \eqref{eq:Condition_for_optimality} behaves as $n \to \infty$. To study the large $n$ behavior we will need to the following bounds on the $ \mathsf{h}_{\nu}$ \cite{segura2011bounds,baricz2015bounds}: for $\nu > \frac{1}{2}$ 
\begin{align}
 \mathsf{h}_{\nu}(x)=   \frac{x}{ \frac{2\nu-1}{2}+\sqrt{ \frac{(2\nu-1)^2}{4} +x^2}} \cdot g_\nu(x),
\end{align}
where 
\begin{align}
1 \ge g_\nu(x) \ge    \frac{ \frac{2\nu-1}{2}+\sqrt{ \frac{(2\nu-1)^2}{4} +x^2}}{ \nu+\sqrt{ \nu^2 +x^2} }.
\end{align} 

Now let $\sfR= c\sqrt{n}$ for some $c>0$.  The goal is to understand the behavior of
\begin{equation}
\bbE \left[      \mathsf{h}_{\frac{n}{2}}^2\left(  \frac{\|  \sqrt{s}\bfZ\| \sfR}{s} \right) +     \mathsf{h}_{\frac{n}{2}}^2\left(  \frac{\|  {\bfx}+\sqrt{s}\bfZ\| \sfR}{s} \right) \right]
\end{equation}
as $n$ goes to infinity. 
First, let 
\begin{align}
V_n= \frac{\| \bfZ\|}{\sqrt{n}},
\end{align}
and note that 
\begin{align}
&\lim_{n \to \infty} \bbE \left[      \mathsf{h}_{\frac{n}{2}}^2\left(  \frac{\|  \sqrt{s}\bfZ\| c \sqrt{n}}{s} \right)  \right]\notag\\
&= \lim_{n \to \infty}  \bbE \left[   \left(  \frac{ \frac{ c V_n }{\sqrt{s}}}{ \frac{n-1}{2n}+\sqrt{ \frac{(n-1)^2}{4n^2} + \left(\frac{ c V_n }{\sqrt{s}} \right)^2}} \cdot g_{ \frac{n}{2}}  \left( \frac{ c V_n }{\sqrt{s}} n \right)\right)^2 \right]\\
&= \bbE \left[    \lim_{n \to \infty}   \left(  \frac{ \frac{ c V_n }{\sqrt{s}}}{ \frac{n-1}{2n}+\sqrt{ \frac{(n-1)^2}{4n^2} + \left(\frac{ c V_n }{\sqrt{s}} \right)^2}} \cdot g_{ \frac{n}{2}}  \left( \frac{ c V_n }{\sqrt{s}} n \right)\right)^2 \right] \label{eq:Applying_DCT} \\
 &= \frac{c^2 }{ \left(  \frac{\sqrt{s}}{2}+\sqrt{ \frac{s}{4} + c^2} \right)^2},\label{eq:using_Law_large_numbers}
\end{align} 
where \eqref{eq:Applying_DCT} follows from the dominated convergence theorem, and \eqref{eq:using_Law_large_numbers} follows since by the law of large numbers we have, almost surely, that 
\begin{align}
\lim_{n \to \infty} V_n^2= \lim_{n \to \infty} \frac{1}{n} \sum_{i=1}^n Z_i^2 =\bbE[Z^2]=1.
\end{align}

Second, let 
\begin{align}
W_n= \frac{\|  {\bfx}+\sqrt{s}\bfZ\|}{\sqrt{n}},
\end{align}
where without loss of generality we take $\bfx=[ \sfR, 0, \ldots, 0]$

\begin{align}
& \lim_{n \to \infty} \bbE \left[         \mathsf{h}_{\frac{n}{2}}^2\left(  \frac{ \|  {\bfx}+\sqrt{s}\bfZ\| c \sqrt{n}}{s} \right) \right] \notag\\
&= \lim_{n \to \infty}  \bbE \left[   \left(  \frac{ \frac{ c W_n }{s}}{ \frac{n-1}{2n}+\sqrt{ \frac{(n-1)^2}{4n^2} + \left(\frac{ c W_n }{s} \right)^2}} \cdot g_{ \frac{n}{2}}  \left( \frac{ cW_n }{s} n \right)\right)^2 \right]\\
&=   \bbE \left[ \lim_{n \to \infty}  \left(  \frac{ \frac{ c W_n }{s}}{ \frac{n-1}{2n}+\sqrt{ \frac{(n-1)^2}{4n^2} + \left(\frac{ c W_n }{s} \right)^2}} \cdot g_{ \frac{n}{2}}  \left( \frac{ cW_n }{s} n \right)\right)^2 \right]  \label{eq:Applying_DCT_v2} \\
&=    \frac{ c^2 (c^2+ s)}{ \left( \frac{s}{2}+\sqrt{ \frac{s^2}{4} +c^2( c^2+ s)  } \right)^2} , \label{eq:using_Law_large_numbers_second}
\end{align} 
where \eqref{eq:Applying_DCT_v2} follows from the dominated convergence theorem and 
where \eqref{eq:using_Law_large_numbers_second} follows since by the strong law of large numbers we have that almost surely
\begin{align}
\lim_{n \to \infty} W_n^2&= \lim_{n \to \infty} \frac{1}{n}   ( \sqrt{s} Z_1+c\sqrt{n})^2 + s \lim_{n \to \infty}  \frac{1}{n} \sum_{i=2}^nZ_i^2\\
 &= c^2+ s.
\end{align}

Combining \eqref{eq:using_Law_large_numbers} and \eqref{eq:using_Law_large_numbers_second} with \eqref{eq:Condition_for_optimality} we arrive at
\begin{align}
& \int_{\sigma_1^2}^{\sigma_2^2} \frac{{ \frac{c^2 }{ \left(  \frac{\sqrt{s}}{2}+\sqrt{ \frac{s}{4} + c^2} \right)^2}} +      \frac{ c^2 (c^2+ s)}{ \left( \frac{s}{2}+\sqrt{ \frac{s^2}{4} +c^2( c^2+ s)  } \right)^2} -1}{s^2} \rmd s =0.
\end{align}

\section{Proof of Theorem~\ref{thm:Main_Results_Scalar}}
\label{Sec:main_result_scalar}
\subsection{KKT Conditions}\label{Sec:KKT_Scalar}
\begin{lemma}\label{lem:KKT_Scalar} $P_{X^\star}$ maximizes \eqref{eq:Secracy_CAP} if and only if 
	\begin{align}
	\Xi(x) &= C_s(\sigma_1^2, \sigma_2^2,\sfR,1), \qquad x \in \supp(P_{X^\star}),\\
	\Xi(x) &\le C_s(\sigma_1^2, \sigma_2^2, \sfR,1), \qquad x \in  [-\sfR,\sfR] ,
	\end{align}
	where  for $x \in \mathbb{R}$
	\begin{align}
	\Xi(x)&=\sfD(f_{Y_1|X}(\cdot|x) \|f_{Y_1^\star})- \sfD(f_{Y_2|X}(\cdot|x) \|f_{Y_2^\star})\\
	&=\bbE \left[ g(Y_1) |X=x \right] +\log\left(\frac{\sigma_2}{\sigma_1}\right), \label{eq:Writing_KKT_as_statistics_Scalar}
	\end{align}
	and where 
	\begin{align} \label{eq:functiong_Scalar}
	g(y)=\bbE\left[\log\frac{f_{Y_2^\star}(y+N)}{f_{Y_1^\star}(y)}\right],  \qquad  y\in \mathbb{R}, 
	\end{align} 
	with $N\sim {\cal N}(0,\sigma_2^2-\sigma_1^2)$. 
\end{lemma} 
\begin{IEEEproof}
	The first part of Lemma~\ref{lem:KKT_Scalar} was shown in \cite{ozel2015gaussian}. The proof of \eqref{eq:Writing_KKT_as_statistics_Scalar} goes as follows: 
	\begin{align}
	&\sfD(f_{Y_1|X}(\cdot|x) \|f_{Y_1^\star})- \sfD(f_{Y_2|X}(\cdot|x) \|f_{Y_2^\star})-\log\left(\frac{\sigma_2}{\sigma_1}\right)\\
	&=\int_{-\infty}^{\infty} \log\frac{1}{f_{Y_1^\star}(y)} \phi_{\sigma_1}(y-x) {\rm d}y \notag\\
	& \quad -\int_{-\infty}^{\infty} \log\frac{1}{f_{Y_2^\star}(y)} \bbE[\phi_{\sigma_1}(y-x-N)] {\rm d}y \label{eq:intro_N} \\
	&=\int_{-\infty}^{\infty} \log\frac{1}{f_{Y_1^\star}(y)} \phi_{\sigma_1}(y-x) {\rm d}y \notag\\
	& \quad -\int_{-\infty}^{\infty} \bbE\left[\log\frac{1}{f_{Y_2^\star}(y+N)}\right] \phi_{\sigma_1}(y-x) {\rm d}y \label{eq:change_var} \\
	&=\int_{-\infty}^{\infty} \bbE\left[\log\frac{f_{Y_2^\star}(y+N)}{f_{Y_1^\star}(y)}\right] \phi_{\sigma_1}(y-x) {\rm d}y\\
	&=\int_{-\infty}^{\infty} g(y) \phi_{\sigma_1}(y-x) {\rm d}y,
	\end{align}
	where in~\eqref{eq:intro_N} we have introduced $N\sim {\cal N}(0,\sigma_2^2-\sigma_1^2)$; and in~\eqref{eq:change_var} we applied the change of variable $y \mapsto y+N$.  This concludes the proof. 
\end{IEEEproof}

{
\subsection{Implicit Upper Bound}
	A consequence of the KKT conditions of Lemma~\ref{lem:KKT_Scalar} is the inclusion
	\begin{equation}
	\supp(P_{X^\star}) 
	\subseteq  \left\{x \in [-\sfR,\sfR] :   \Xi(x) - C_s=0 \right \}   \label{eq:InclusiongInequality}
	\end{equation} 
	which suggests the following upper bound on the number of support points of $P_{X^\star}$:
	\begin{align}
	&|\supp(P_{X^\star})|  \notag\\
	&\le \rmN\left([-\sfR,\sfR],   \Xi(x) - C_s(\sigma_1^2, \sigma_2^2, \sfR,1) \right) \label{eq:Zeros_Inclusion_Bound} \\
	&= \rmN\left([-\sfR,\sfR],\bbE \left[ g(Y_1) +\log\left(\frac{\sigma_2}{\sigma_1}\right)-C_s \Big| X=x \right] \right)  \label{eq:Using_Def_of_Xi} \\
	&\le \scrS\left( g(\cdot)+\log\left(\frac{\sigma_2}{\sigma_1}\right)-C_s \right) \label{eq:applying_Oscillationthm}  \\
	&\le \rmN\left(\mathbb{R}, g(\cdot)+\log\left(\frac{\sigma_2}{\sigma_1}\right)-C_s\right) \\
	&= \rmN\left([-L,L], g(\cdot)+\log\left(\frac{\sigma_2}{\sigma_1}\right)-C_s\right) \label{eq:Lemmaboundedsupport} \\
	&< \infty, \label{eq:Follows_by_analyticity}
	\end{align}
	where \eqref{eq:Using_Def_of_Xi} follows from using \eqref{eq:Writing_KKT_as_statistics_Scalar};   \eqref{eq:applying_Oscillationthm} follows from applying Karlin's oscillation Theorem~\ref{thm:OscillationThoerem} and the fact that the Gaussian pdf is   a strictly totally positive kernel, which was shown in \cite{karlin1957polya};  \eqref{eq:Lemmaboundedsupport} is proved in Lemma~\ref{Lem:boundedsupport} in the Appendix;  and \eqref{eq:Follows_by_analyticity}   follows because $g(\cdot)$ is an analytic function in $(-L,L)$. The implicit upper bound~\eqref{eq:Implicit_Upper_Bound_Scalar} of Theorem~\ref{thm:Main_Results_Scalar} follows from~\eqref{eq:Lemmaboundedsupport} and~\eqref{eq:Follows_by_analyticity}.
}

	\subsection{Explicit Upper Bound}
	The key to finding an explicit upper bound on the number of zeros will be the following  complex-analytic result. 
	\begin{lemma}[Tijdeman's Number of Zeros Lemma \cite{Tijdeman1971number}]\label{lem:number of zeros of analytic function}
		Let $L, s, t$ be positive numbers such that $s>1$. For the complex valued function $f\neq  0$ which is analytic on $|z|<(st+s+t)L$, its number of zeros $  \rmN(\cD_L,f)$ within the disk $\cD_L = \{z\colon |z|\le L\} $ satisfies
		\begin{align}
		& \rmN(\cD_L,f) \notag\\
		& \le \frac{1}{\log s} \left(\log \max_{|z|\le (st+s+t)L } |f(z)|   -\log \max_{|z|\le tL} |f(z)|\right) \text{.} \label{eq:Tijdeman}
		\end{align}
		\end{lemma}

Furthermore, the following loosened version of the implicit upper bound in \eqref{eq:Implicit_Upper_Bound_Scalar} will be useful. 
		\begin{lemma}
			\begin{align}
			|\supp(P_{X^\star})| 
			&\le \rmN\left([-L,L], h (\cdot) \right) +1 \label{eq:productfy1}
			\end{align}
			where 
			\begin{align}
			&\frac{h(y)}{ \sigma_1^2 f_{Y_1}(y)}  \notag\\
			&=  \frac{ \bbE_N \left[  \bbE[X^\star| Y_2=y+N] \right] -y}{\sigma_2^2}-  \frac{\bbE[X^\star| Y_1=y] -y}{\sigma_1^2}  \label{eq:First_representation_h}\\
			&= \frac{\bbE\left[N\log f_{Y_2}(y+N) \right]}{\sigma^2_2-\sigma^2_1} -  \frac{\bbE[X^\star| Y_1=y] -y}{\sigma_1^2}, 
			\end{align}
			and where $N\sim {\cal N}(0,\sigma_2^2-\sigma_1^2)$. 
		\end{lemma}
		\begin{IEEEproof}
			Starting from~\eqref{eq:Lemmaboundedsupport}, we can write
			\begin{align}
			|\supp(P_{X^\star})| &\le \rmN\left([-L,L], g(\cdot)+\log\left(\frac{\sigma_2}{\sigma_1}\right)-C_s\right) \\
			&\le \rmN\left([-L,L], g'(\cdot)\right) +1 \label{eq:Rolle1}\\
			&=\rmN\left([-L,L], \sigma_1^2 f_{Y_1}(\cdot) g'(\cdot)\right) +1 \label{eq:productfy3}
			\end{align}
			where in step \eqref{eq:Rolle1} we have applied Rolle's theorem, and in step \eqref{eq:productfy3} we used the fact that multiplying by a strictly positive function (i.e., $\sigma_1^2 f_{Y_1}$) does not change the number of zeros. The first derivative of $g$ can be computed as follows:
			\begin{align}
			g'(y) &= \bbE\left[\frac{\rm d}{{\rm d}y}\log f_{Y_2}(y+N) \right] -\frac{ \rm d }{ {\rm d}y} \log f_{Y_1}(y) \label{eq:der11}\\
			&=  \frac{ \bbE_N \left[  \bbE[X^\star| Y_2=y+N] \right] -y}{\sigma_2^2}-  \frac{\bbE[X^\star| Y_1=y] -y}{\sigma_1^2},
			\end{align}
			where in the last step we have used the well-known Tweedy's formula (see for example \cite{esposito1968relation,dytso2020general}):
			\begin{equation}
			\bbE[X^\star| Y_i=y] = y  +\sigma^2_i\frac{ \rm d }{ {\rm d}y}\log f_{Y_i}(y).  
			\end{equation}  
			An alternative expression for the first term in the right-hand side (RHS) of~\eqref{eq:der11} is as follows:
			\begin{align}
			&\bbE\left[\frac{\rm d}{{\rm d}y}\log f_{Y_2}(y+N) \right] \notag\\
			&= \int_{-\infty}^{\infty} f_N(n)\frac{ \rm d }{ {\rm d}y}\log f_{Y_2}(y+n) {\rm d}n \\
			&=-\int_{-\infty}^{\infty} \left(\frac{ \rm d }{ {\rm d}n}f_N(n)\right)\cdot \log f_{Y_2}(y+n) {\rm d}n \\
			&=\int_{-\infty}^{\infty} \frac{n }{\sigma^2_2-\sigma^2_1}f_N(n)\cdot \log f_{Y_2}(y+n) {\rm d}n \\
			&=\frac{1}{\sigma^2_2-\sigma^2_1} \bbE\left[N\log f_{Y_2}(y+N) \right],
			\end{align}
			where $f_N(n)=\phi_{\sqrt{\sigma^2_2-\sigma^2_1}}(n)$.
			The proof is concluded by letting 
			\begin{align} \label{eq:productfy2}
			h(y)\triangleq \sigma_1^2 f_{Y_1}(y)g'(y).
			\end{align}
		
		\end{IEEEproof}

			To apply Tijdeman's number of zeros Lemma, upper and lower bounds to the maximum module of the complex analytic extension of $h$ over the disk ${\cal D}_L = \{z: |z|\le L \}$ are proposed in Lemma~\ref{lem:moduls_upper_bound} and Lemma~\ref{lem:moduls_lower_bound} in the Appendix. Using those bounds, we can provide an upper bound on the number of mass points as follows:
			\begin{align}
			&\rmN\left([-  L,  L],h(\cdot)\right)\\
			&\le \rmN\left( \mathcal{D}_{L},\breve{h}(\cdot) \right) \label{eq:Extension_to_complex_bound}\\
			&\le  \min_{s> 1,\, t > 0 } \left\{ \frac{\log  \frac{\max_{|z| \le (st+s+t)L}|\breve h (z)| }{\max_{|z| \le t L} |\breve h (z)|}}{\log s}  \right\} \label{frm:lem: bd on the no of zeros of analytic func} \\
			&\le  \log  \frac{\frac{  \rme^{ \frac{(2\rme+1)^2L^2}{2\sigma_1^2} } }{\sqrt{2\pi\sigma_1^2}} \left( a_1 (2\rme+1)^2L^2 +a_2(2\rme+1)L+a_3 \right)
			}{ \left( c_1 L   - c_2 \sfR     \right) 
			\frac{  \exp\left(  -\frac{(L+\sfR)^2}{2\sigma_1^2} \hspace{-0.1cm}\right) }{\sqrt{2\pi \sigma_1^2}} } \label{eq:choosing_s_and_t}\\
		&= \frac{(2\rme+1)^2L^2}{2\sigma_1^2}+ \frac{(L+\sfR)^2}{2\sigma_1^2}   \notag\\
		& \quad +   \log  \frac{  a_1 (2\rme+1)^2L^2 +a_2(2\rme+1)L+a_3 
		}{  c_1 L   - c_2 \sfR      } \\
		&= \frac{(2\rme+1)^2(d_1 \sfR +d_2)^2}{2\sigma_1^2}+ \frac{( (d_1+1)\sfR+d_2)^2}{2\sigma_1^2}   \notag\\
		&  +   \log  \frac{  a_1 (2\rme+1)^2(d_1 \sfR +d_2)^2 +a_2(2\rme+1)(d_1 \sfR +d_2)+a_3 
		}{ (c_1d_1-c_2) \sfR +c_1 d_2   }  \label{eq:inserting_R_def}\\
		& \le   b_1  \frac{\sfR^2}{\sigma^2_1}+ b_2 +\log \frac{b_3\sfR^2+b_4 \sfR+b_5}{ b_6 \sfR+b_7} \label{eq:square_bound}\\
		& \le   b_1  \frac{\sfR^2}{\sigma^2_1}+ O( \log(\sfR) ) ,  \label{eq:Big_O_bound}
		\end{align}
		where \eqref{eq:Extension_to_complex_bound} follows because extending to larger domain can only increase the number of zeros; \eqref{frm:lem: bd on the no of zeros of analytic func} follows from the Tijdeman's Number of Zeros Lemma;      \eqref{eq:choosing_s_and_t} follows from choosing $s=\rme$ and $t=1$ and using bounds in Lemma~\ref{lem:moduls_upper_bound} and Lemma~\ref{lem:moduls_lower_bound}; \eqref{eq:inserting_R_def} follows from using the value of $L$ in \eqref{eq:Bound_on_R}; \eqref{eq:square_bound} using the bound $(a+b)^2 \le 2 (a^2+b^2)$ and defining 
		\begin{subequations}
		\begin{align}
		b_1&= (2\rme+1)^2d_1^2+(d_1+1)^2\\
		&=  (2\rme+1)^2 \left( \frac{\sigma_2+\sigma_1}{ \sigma_2-\sigma_1} \right)^2+ \left(\frac{\sigma_2+\sigma_1}{ \sigma_2-\sigma_1}+1 \right)^2  \\
		b_2&= \frac{((2\rme+1)^2+1)d_2^2}{\sigma_1^2} \\
		&= \frac{((2\rme+1)^2+1)}{\sigma_1^2}   \frac{ \frac{\sigma_2^2-\sigma_1^2}{\sigma_2^2}+2C_s}{ \frac{1}{\sigma_1^2}-\frac{1}{\sigma_2^2} } \\
		&=((2\rme+1)^2+1) \left (1+ 2\frac{\sigma_2^2}{\sigma_2^2-\sigma_1^2} C_s \right)\\
		b_3&=2 (2 \rme+1)^2 a_1 d_1^2\\
		&=    2 (2 \rme+1)^2 \frac{ 3 \sigma_1^2}{ \sigma_2^2 \sqrt{\sigma_2^2-\sigma_1^2}} \left(\frac{\sigma_2+\sigma_1}{ \sigma_2-\sigma_ 1} \right)^2\\
		b_4&=(2 \rme+1) d_1  a_2\\
		&=  (2 \rme+1)  \frac{\sigma_2+\sigma_1}{ \sigma_2-\sigma_ 1} \left( \frac{ \sqrt{2} \sigma_1^2}{  \sqrt{ \sigma_2^2} \sqrt{\sigma_2^2-\sigma_1^2}} +2 \right) \\
		b_5&=2(2\rme+1)^2 a_1d_2^2+  (2\rme +1) a_2 d_2 +a_3\\
		&=2(2\rme+1)^2  \frac{ 3 \sigma_1^2}{ \sigma_2^2 \sqrt{\sigma_2^2-\sigma_1^2}} \left(  \frac{ \frac{\sigma_2^2-\sigma_1^2}{\sigma_2^2}+2C_s}{ \frac{1}{\sigma_1^2}-\frac{1}{\sigma_2^2} } \right) \notag\\
		&+  (2\rme +1) \left(\frac{ \sqrt{2} \sigma_1^2}{  \sqrt{ \sigma_2^2} \sqrt{\sigma_2^2-\sigma_1^2}} +2 \right) \sqrt{ \frac{ \frac{\sigma_2^2-\sigma_1^2}{\sigma_2^2}+2C_s}{ \frac{1}{\sigma_1^2}-\frac{1}{\sigma_2^2} } }   \notag\\
		& +\frac{\sigma_1^2}{\sqrt{\sigma_2^2-\sigma_1^2}} \cdot  \sqrt{  |\log(2\pi\sigma^2_2)|^2 +  \frac{ 24  (\sigma_2^2-\sigma_1^2)^2 }{\sigma^4_2}  + \pi^2}   \\
		b_6&= c_1d_1-c_2\\
		&= \frac{\sigma_2^2-\sigma_1^2 }{\sigma_2^2} \frac{\sigma_2+\sigma_1}{ \sigma_2-\sigma_ 1}  -  \frac{\sigma_2^2+\sigma_1^2 }{\sigma_2^2} =2 \frac{\sigma_1}{\sigma_2}\\
		b_7&=c_1 d_2\\
		&=  \frac{\sigma_2^2-\sigma_1^2 }{\sigma_2^2}\sqrt{ \frac{ \frac{\sigma_2^2-\sigma_1^2}{\sigma_2^2}+2C_s}{ \frac{1}{\sigma_1^2}-\frac{1}{\sigma_2^2} } };
		\end{align}
		\end{subequations}
		  and  \eqref{eq:Big_O_bound} follows from the fact that the  $b_1,b_3,b_4$ and $b_6$ coefficients  do not depend on $\sfR$ and the fact that the coefficients  $b_2,b_5$ and $b_4$, while do depend on $\sfR$ through $C_s$, do not grow with~$\sfR$.  The fact that $C_s$ does not grow with $\sfR$ follows from the bound in \eqref{eq:C_G}.

		Finally, the explicit upper bound on the number of support points of $P_{X^\star}$ in~\eqref{eq:Upper_Bound_Explicit_Scalar} is a consequence of~\eqref{eq:Big_O_bound}.

\section{Proof of Theorem~\ref{thm:Capacitiy_Small}} 
\label{sec:thm:Capacitiy_Small}
Using the KKT conditions in \eqref{eq:NewOptimalityEquations}, we have that for $\bfx=[\sfR, 0, \ldots, 0]$ 
\begin{align}
C_s(\sigma_1^2, \sigma_2^2, \sfR,n)&=\Xi(\bfx;P_{\bfX_{\sfR}})\\
&=\sfD(f_{\bfY_1|\bfX}(\cdot| \bfx) \|f_{\bfY_1^\star})- \sfD(f_{\bfY_2|\bfX}(\cdot|\bfx) \|f_{\bfY_2^\star})\\
&=  \frac{1}{2} \int_{\sigma_1^2}^{\sigma_2^2} \frac{\sfR^2 -\sfR^2\bbE \left[      \mathsf{h}_{\frac{n}{2}}^2\left(  \frac{\|  \sfR+\sqrt{s}\bfZ\| \sfR}{s} \right) \right] }{s^2} \rmd s
\end{align}
where the last expression was computed in \eqref{eq:diff_of_KL_s}. This concludes the proof. 

\section{Conclusion} \label{sec:conclusion}
This paper focuses on the secrecy-capacity of the  $n$-dimensional vector Gaussian wiretap channel under the peak-power (or amplitude constraint) in a so-called low (but not vanishing) amplitude regime. In this regime, the optimal input distribution $P_{\bfX_\sfR}$ is supported on a single  $n$-dimensional sphere of radius $\mathsf{R}$. The paper has identified the largest $\bar{\mathsf{R}}_n$ such that the distribution $P_{\bfX_\sfR}$ is optimal. In addition, the asymptotic of $\bar{\mathsf{R}}_n$ has been completely characterized as dimension $n$ approaches infinity. As a by-product of the analysis, the capacity in the low amplitude regime has also been characterized in more or less closed-form. The paper has also provided a number of supporting numerical examples. Implicit and explicit upper bounds have been proposed on the number of mass points for the optimal input distribution $P_{X^\star}$ in the scalar case with $n=1$. As part of ongoing work, we are trying to resolve the conjecture that was made regarding the number of zeros of the function defined through the ratios of Bessel functions.

There are several interesting future directions. For example, one interesting direction would be to determine a regime in which a mixture of a mass point at zero and $P_{\bfX_\sfR}$ is optimal. It would also be interesting to establish a lower bound on the number of mass points in the support of the optimal input distribution when $n=1$. We note that  such a lower bound was obtained for a point-to-point channel in \cite{DytsoAmplitute2020}. We finally remark that the extension of the results of this paper to nondegraded wiretap channels is not trivial and also constitutes and interesting but ambitious future direction.

\begin{appendices} 

\section{Examples of the Function $G_{\sigma_1,\sigma_2,\sfR,n}$ }
 
\begin{figure}

	\centering
	\subfloat[$n=3$, $\sigma_1=1$ and $\sigma_2=2$. ]{\input{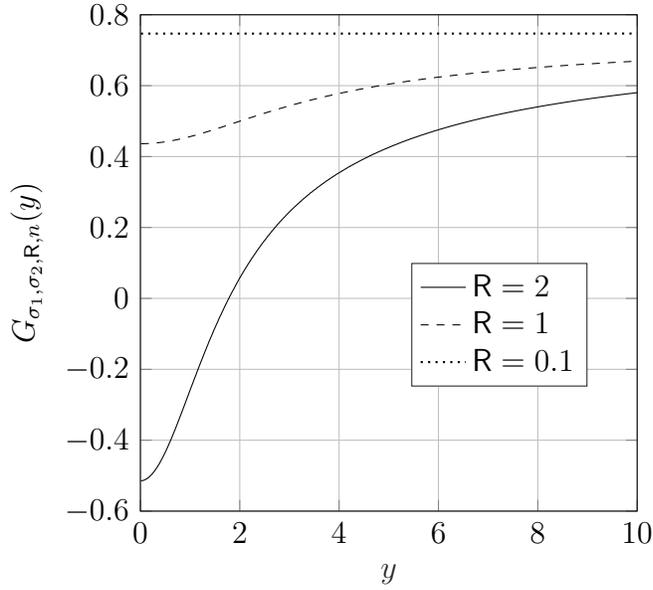}}
	\subfloat[$n=11$, $\sigma_1=1$ and $\sigma_2=2$. ]{\input{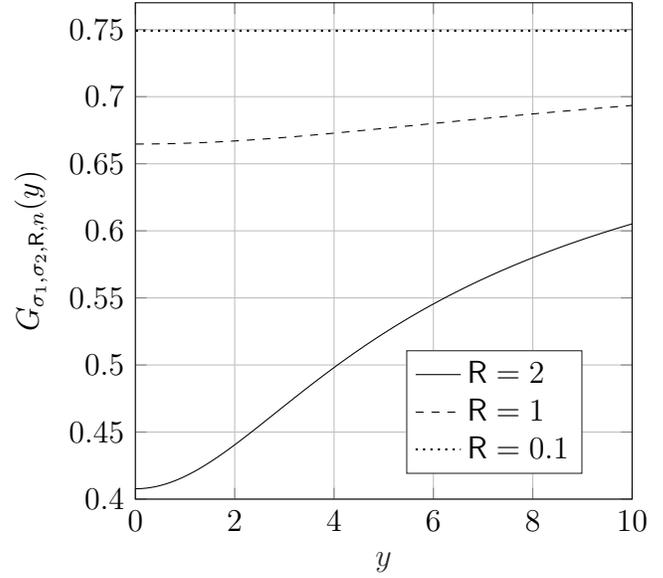}}

~
	\subfloat[$n=4$, $\sigma_1=3$ and $\sigma_2=3.1$. ]
	{\input{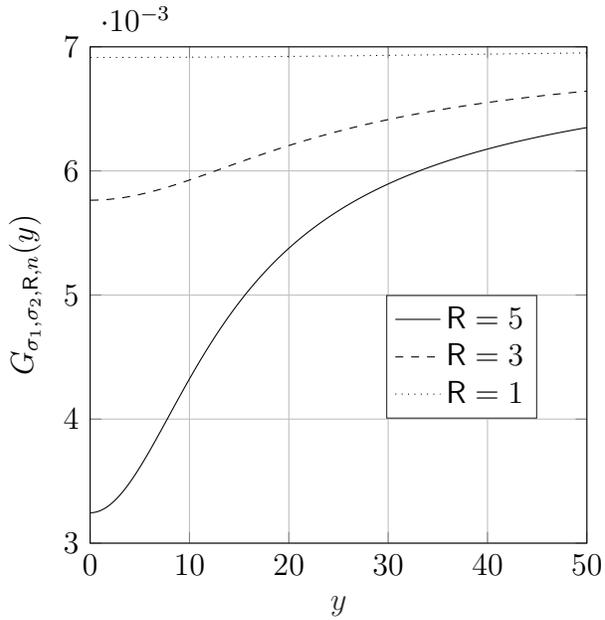}}
	\subfloat[$n=11$, $\sigma_1=3$ and $\sigma_2=3.1$. ]
{\input{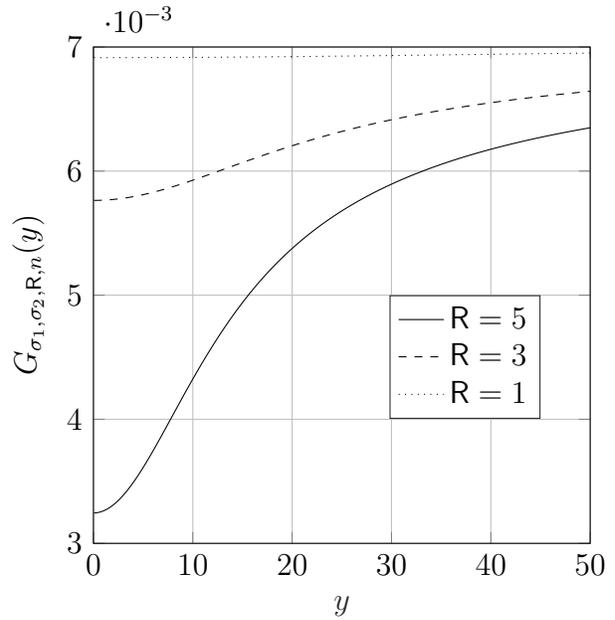}}	

	\caption{Examples of the function $G_{\sigma_1,\sigma_2,\sfR,n}$ defined in \eqref{eq:Definition_of_G_function}.   }
	\label{fig:Examples_G}
\end{figure}

In this section, we give supporting numerical arguments that the function $G_{\sigma_1,\sigma_2,\sfR,n}$ defined in~\eqref{eq:Definition_of_G_function} has at most one sign change. 
Figure~\ref{fig:Examples_G} demonstrates the behavior of the function  $G_{\sigma_1,\sigma_2,\sfR,n}$. In addition, the code that generates the function $G_{\sigma_1,\sigma_2,\sfR,n}$ for various values of $n, \sigma_1$ and $\sigma_2$ is provided in~\cite{GithubData}.

\label{app:Examples_G_func}
\section{Derivative of the Secrecy-Density}
\begin{lemma} \label{Lemma:derivative_Xi}
	The derivative of the secrecy-density for the input $P_{\bfX_{\sfR}}$ is
	\begin{align} \label{eq:derivative_Xi_lemma}
	\Xi'(\|\bfx\|;P_{\bfX_{\sfR}}) =\|\bfx \|\:\bbE\left[\widetilde{M}_2(\sigma_1 Q_{n+2})-M_1(\sigma_1 Q_{n+2}) \right]
	\end{align}
	where $Q_{n+2}^2$ is a noncentral chi-square random variable with $n+2$ degrees of freedom and noncentrality parameter $\frac{\|\bfx\|^2}{\sigma_1^2}$ and
	\begin{align}
	M_i(y) &= \frac{1}{\sigma_i^2}\left(\frac{\sfR}{y}\sfh_{  \frac{n}{2} }\left(\frac{\sfR}{\sigma_i^2}y\right)-1\right), \qquad i\in \{1,2\} \\
	\widetilde{M}_2(y) &= \bbE\left[M_2(\|y+\bfW \|)\right],
	\end{align}
	where $\bfW \sim {\cal N}(\mathbf{0}_{n+2},(\sigma_2^2-\sigma_1^2)\bfI_{n+2})$.
\end{lemma}
\begin{proof}
	We start with the secrecy-density expressed in spherical coordinates. A quick way to get to the information densities in this coordinate system is to note that:
	\begin{align}
	&I(\bfX;\bfY_i) \nonumber\\
	 &= h(\bfY_i)-h(\bfN_i) \\
	&=h(\|\bfY_i \|)+(n-1)\bbE[\log\|\bfY_i \|]+h_\lambda\left(\frac{\bfY_i}{\|\bfY_i \|}  \right)- h(\bfN_i) \label{eq:spherical} \\
	&=h(\|\bfY_i \|^2)+\left(\frac{n}{2}-1\right)\bbE[\log\|\bfY_i \|^2] \nonumber\\
	&\quad+\log\frac{\pi^{\frac{n}{2}}}{\Gamma\left(\frac{n}{2}\right)}- \frac{n}{2}\log(2\pi e \sigma_i^2) \label{eq:square} \\
	&=h\left(\sigma_i^2 \left\|\frac{\bfX}{\sigma_i}+\widetilde{\bfN}_i  \right\|^2\right) \nonumber\\
	&\quad+\left(\frac{n}{2}-1\right)\bbE\left[\log\left(\sigma_i^2\left\|\frac{\bfX}{\sigma_i}+\widetilde{\bfN}_i  \right\|^2\right)\right] \nonumber\\
	&\quad+\log\frac{\pi^{\frac{n}{2}}}{\Gamma\left(\frac{n}{2}\right)}- \frac{n}{2}\log(2\pi e \sigma_i^2) \label{eq:chi_square_introduction} \\
	&=h\left( \left\|\frac{\bfX}{\sigma_i}+\widetilde{\bfN}_i  \right\|^2\right)+\left(\frac{n}{2}-1\right)\bbE\left[\log\left\|\frac{\bfX}{\sigma_i}+\widetilde{\bfN}_i  \right\|^2\right] \nonumber\\
	&\quad- \log\left((2 e)^\frac{n}{2}  \Gamma\left(\frac{n}{2}\right)\right),
	\end{align}
	where~\eqref{eq:spherical} holds by~\cite[Lemma~6.17]{lapidoth2003capacity} and by independence between $\|\bfY_i \|$ and $\frac{\bfY_i}{\|\bfY_i \|}$; the term $h_\lambda(\cdot)$ is a differential entropy-like quantity for random vectors on the $n$-dimensional unit sphere~\cite[Lemma~6.16]{lapidoth2003capacity}; \eqref{eq:square} holds because $\frac{\bfY_i}{\|\bfY_i \|}$ is uniform on the unit sphere and thanks to~\cite[Lemma~6.15]{lapidoth2003capacity}; the term $\Gamma(z)$ is the gamma function; and in~\eqref{eq:chi_square_introduction} we have $\widetilde{\bfN}_i\sim {\cal N}(\mathbf{0}_n,\bfI_n)$. It is now immediate to write the secrecy-density as follows:
	\begin{equation} \label{eq:secrecy-density_norm}
	\Xi(\|\bfx \|;P_{\bfX}) = i_1(\|\bfx \|;P_{\bfX})-i_2(\|\bfx \|;P_{\bfX})
	\end{equation}
	where
	\begin{align}
	&i_j(\|\bfx \|;P_{\bfX}) \nonumber\\
	&= -\int_0^{\infty} f_{\chi^2_{n}(\frac{\|\bfx \|^2}{\sigma_j^2})}(y) \log \frac{\int_0^\sfR f_{\chi^2_{n}(\frac{t^2}{\sigma_j^2})}(y)dP_{\|\bfX\|}(t)}{y^{\frac{n}{2}-1}}dy \nonumber\\
	&\quad - \log\left((2e)^{\frac{n}{2}} \Gamma\left( \frac{n}{2} \right) \right),
	\end{align}
	for $j\in\{1,2\}$. The term $f_{\chi^2_n(\lambda)}(y)$ is the noncentral chi-square pdf with $n$ degrees of freedom and noncentrality parameter $\lambda$.

	Given two values $\rho_1,\rho_2$ with $\rho_1>\rho_2$, write
	\begin{align}
	&i_j(\rho_1;P_{\bfX})-i_j(\rho_2;P_{\bfX}) \nonumber\\
	 &= \int_{0}^{\infty} \left(f_{\chi^2_{n}(\frac{\rho_1^2}{\sigma_j^2})}(y)-f_{\chi^2_{n}(\frac{\rho_2^2}{\sigma_j^2})}(y)\right)\log\frac{y^{\frac{n}{2}-1}}{f_{\|\frac{\bfY}{\sigma_j}\|^2}(y;P_{\bfX})} dy \\
	&=\int_{0}^{\infty} \left(F_{\chi^2_{n}(\frac{\rho_2^2}{\sigma_j^2})}(y)-F_{\chi^2_{n}(\frac{\rho_1^2}{\sigma_j^2})}(y)\right)\frac{d}{dy}\log\frac{y^{\frac{n}{2}-1}}{f_{\|\frac{\bfY}{\sigma_j}\|^2}(y;P_{\bfX})} dy \label{eq:der1}
	\end{align}
	where we have integrated by parts and where $F_{\chi^2_{n}(\lambda)}(y)$ is the cumulative distribution function of $\chi^2_{n}(\lambda)$. Now notice that
	\begin{equation}
	\int_{0}^{\infty} \left(F_{\chi^2_{n}(\frac{\rho_2^2}{\sigma_j^2})}(y)-F_{\chi^2_{n}(\frac{\rho_1^2}{\sigma_j^2})}(y)\right) dy = \frac{\rho_1^2 - \rho_2^2}{\sigma_j^2}. \label{eq:auxpdf}
	\end{equation}
	Since $\chi^2_{n}(\frac{\rho_1^2}{\sigma_j^2})$ statistically dominates $\chi^2_{n}(\frac{\rho_2^2}{\sigma_j^2})$, the integrand function in \eqref{eq:auxpdf} is always positive. We can introduce an auxiliary output random variable $Q_j$, for $j\in\{1,2\}$, with pdf
	\begin{equation}\label{eq:def_fQ}
	f_{Q_j}(y;\rho_1,\rho_2) =\frac{\sigma_j^2}{\rho_1^2 - \rho_2^2} \left(F_{\chi^2_{n}(\frac{\rho_2^2}{\sigma_j^2})}(y)-F_{\chi^2_{n}(\frac{\rho_1^2}{\sigma_j^2})}(y)\right), 
	\end{equation} 
	for $y>0$,	to rewrite \eqref{eq:der1} as follows:
	\begin{align}
	&i_j(\rho_1;P_{\bfX})-i_j(\rho_2;P_{\bfX}) \nonumber\\
	 &= -\frac{\rho_1^2-\rho_2^2}{\sigma_j^2}\int_{0}^{\infty} f_{Q_j}(y;\rho_1,\rho_2)\frac{d}{dy}\log\frac{f_{\|\frac{\bfY}{\sigma_j}\|^2}(y;P_{\bfX})}{y^{\frac{n}{2}-1}} dy.\label{eq:afterQ}
	\end{align}
	We evaluate the derivative in \eqref{eq:afterQ} as:
	\begin{align}
	&\frac{d}{dy}\log\frac{f_{\|\frac{\bfY}{\sigma_j}\|^2}(y;P_{\bfX})}{y^{\frac{n}{2}-1}} \nonumber\\
	&= \frac{y^{\frac{n}{2}-1}}{f_{\|\frac{\bfY}{\sigma_j}\|^2}(y;P_{\bfX})}\int_{0}^{\sfR} \frac{d}{dy} \frac{f_{\chi^2_{n}(\frac{t^2}{\sigma_j^2})}(y)}{y^{\frac{n}{2}-1}} dP_{\|\bfX\|}(t) \label{eq:fy} \\
	&=\frac{y^{\frac{n}{2}-1}}{f_{\|\frac{\bfY}{\sigma_j}\|^2}(y;P_{\bfX})}\nonumber\\
	&\quad\int_{0}^{\sfR} \left(\frac{f_{\chi^2_{n-2}(\frac{t^2}{\sigma_j^2})}(y)}{2y^{\frac{n}{2}-1}}-\left(\frac{1}{2}+\frac{\frac{n}{2}-1}{y}\right)\frac{f_{\chi^2_{n}(\frac{t^2}{\sigma_j^2})}(y)}{y^{\frac{n}{2}-1}}\right) dP_{\|\bfX\|}(t) \label{eq:der_chisquare} \\
	&=\bbE\left[\frac{1}{2} \frac{f_{\chi^2_{n-2}(\frac{\|\bfX\|^2}{\sigma_j^2})}(\frac{\left\|\bfY\right\|^2}{\sigma_j^2})}{f_{\chi^2_{n}(\frac{\|\bfX\|^2}{\sigma_j^2})}(\frac{\left\|\bfY\right\|^2}{\sigma_j^2})}-\left(\frac{1}{2}+\frac{\frac{n}{2}-1}{\frac{\left\|\bfY\right\|^2}{\sigma_j^2}}\right) | \frac{\left\|\bfY\right\|^2}{\sigma_j^2}=y\right] \\
	&=\bbE\left[\frac{1}{2}\frac{\|\bfX\|}{\|\bfY\|} \frac{\sfI_{\frac{n}{2}-2}(\frac{\|\bfX\| \|\bfY\|}{\sigma_j^2})}{\sfI_{\frac{n}{2}-1}(\frac{\|\bfX\| \|\bfY\|}{\sigma_j^2})}-\left(\frac{1}{2}+\frac{\frac{n}{2}-1}{\frac{\left\|\bfY\right\|^2}{\sigma_j^2}}\right) | \frac{\left\|\bfY\right\|^2}{\sigma_j^2}=y\right] \label{eq:derlog} \\
	&=\bbE\left[\frac{1}{2}\frac{\|\bfX\|}{\|\bfY\|}\sfh_{\frac{n}{2}}\left(\frac{\|\bfX\| \|\bfY\|}{\sigma_j^2}\right)-\frac{1}{2} |   \frac{\left\|\bfY\right\|^2}{\sigma_j^2}=y\right] \label{eq:recurrence_relation}
	\end{align} 
	where in \eqref{eq:fy} we used
	\begin{equation}
	f_{\|\frac{\bfY}{\sigma_j}\|^2}(y;P_{\bfX}) = \int_{0}^{\sfR} f_{\chi^2_{n}(\frac{t^2}{\sigma_j^2})}(y) dP_{\|\bfX\|}(t);
	\end{equation} in \eqref{eq:der_chisquare} we used the relationship
	\begin{equation} \label{eq:derivative_chisquared}
	\frac{d}{dy} f_{\chi^2_{n}(\rho^2)}(y) = \frac{1}{2} f_{\chi^2_{n-2}(\rho^2)}(y)-\frac{1}{2} f_{\chi^2_{n}(\rho^2)}(y);
	\end{equation}
	and \eqref{eq:recurrence_relation} follows from the recurrence relationship
	\begin{equation}
	\sfI_{\nu-1}(z)-\sfI_{\nu+1}(z)=\frac{2\nu}{z}\sfI_\nu(z).
	\end{equation}
	Putting together \eqref{eq:afterQ} and \eqref{eq:recurrence_relation} we get
	\begin{align}
	&i_j(\rho_1;P_{\bfX})-i_j(\rho_2;P_{\bfX}) \\
	 &= -\frac{\rho_1^2-\rho_2^2}{2\sigma_j^2} \bbE\left[\bbE\left[\frac{\|\bfX\|}{\|\bfY\|}\sfh_{\frac{n}{2}}\left(\frac{\|\bfX\| \|\bfY\|}{\sigma_j^2}\right)-1| \frac{\left\|\bfY\right\|^2}{\sigma_j^2}=Q_j\right]\right].
	\end{align}
	We are now in the position to compute the derivative of the information density as:
	\begin{align}
	&i_j'(\rho;P_{\bfX}) \nonumber\\
	 &= \lim_{h\rightarrow 0} \frac{i_j(\rho+h;P_{\bfX})-i_j(\rho;P_{\bfX})}{h} \\
	&=-\frac{\rho}{\sigma_j^2}\: \bbE\left[\bbE\left[\frac{\|\bfX\|}{\|\bfY\|}\sfh_{\frac{n}{2}}\left(\frac{\|\bfX\| \|\bfY\|}{\sigma_j^2}\right)-1 | \frac{\left\|\bfY\right\|^2}{\sigma_j^2}=Q'\right]\right]
	\end{align}
	where $Q'\sim \chi^2_{n+2}(\frac{\rho^2}{\sigma_j^2})$ thanks to Lemma \ref{Lemma:fQj}.
	
	The final result is obtained by letting
	\begin{align}
	\Xi'(\|\bfx \|;P_{\bfX}) = i_1'(\|\bfx \|;P_{\bfX})-i_2'(\|\bfx \|;P_{\bfX})
	\end{align}
	and by specializing the result to the input $P_{\bfX_{\sfR}}$.
\end{proof}
\begin{lemma}\label{Lemma:fQj}
	Consider the pdf $f_{Q_j}(y;\rho_1,\rho_2)$ defined in~\eqref{eq:def_fQ}. For any $\rho\ge 0$ we have
	\begin{equation}
	\lim_{h\rightarrow 0} f_{Q_j}(y;\rho+h,\rho) = f_{\chi^2_{n+2}(\frac{\rho^2}{\sigma_j^2})}(y), \qquad y>0.
	\end{equation}
\end{lemma}
\begin{proof}
	Thanks to the definition \eqref{eq:def_fQ}, we have
	\begin{align}
	&\lim_{h\rightarrow 0} f_{Q_j}(y;\rho+h,\rho) \nonumber\\
	 &= \lim_{h\rightarrow 0} \frac{\sigma_j^2}{h(2\rho+h)}\left(F_{\chi^2_{n}(\frac{\rho^2}{\sigma_j^2})}(y)-F_{\chi^2_{n}(\frac{(\rho+h)^2}{\sigma_j^2})}(y)\right) \\
	&=\lim_{h\rightarrow 0} \frac{\sigma_j^2}{h(2\rho+h)} \int_{0}^{y} \left(f_{\chi^2_{n}(\frac{\rho^2}{\sigma_j^2})}(t)-f_{\chi^2_{n}(\frac{(\rho+h)^2}{\sigma_j^2})}(t)  \right) dt \\
	&= \frac{\sigma_j^2}{2\rho}\int_{0}^{y} \sum_{i=0}^{\infty} \lim_{h\rightarrow 0} \frac{1}{h} \nonumber\\
	&\quad\left(\frac{\rme^{-\frac{\rho^2}{2\sigma_j^2}}\left(\frac{\rho^2}{2\sigma_j^2}\right)^i}{i!}-\frac{\rme^{-\frac{(\rho+h)^2}{2\sigma_j^2}}\left(\frac{(\rho+h)^2}{2\sigma_j^2}\right)^i}{i!} \right) f_{\chi^2_{n+2i}}(t) dt \label{eq:Poisson_representation} \\
	&= \frac{\sigma_j^2}{2\rho}\int_{0}^{y} \sum_{i=0}^{\infty} \frac{d}{d\rho} \left(\frac{\rme^{-\frac{\rho^2}{2\sigma_j^2}}\left(\frac{\rho^2}{2\sigma_j^2}\right)^i}{i!}\right) f_{\chi^2_{n+2i}}(t) dt \\
	&= \frac{1}{2}\int_{0}^{y} \sum_{i=0}^{\infty}  \left(-\frac{\rme^{-\frac{\rho^2}{2\sigma_j^2}}\left(\frac{\rho^2}{2\sigma_j^2}\right)^i}{i!}\right.\nonumber\\
	&\qquad\qquad\left. +\frac{\rme^{-\frac{\rho^2}{2\sigma_j^2}}\left(\frac{\rho^2}{2\sigma_j^2}\right)^{i-1}}{(i-1)!}{1}(i\ge 1)\right) f_{\chi^2_{n+2i}}(t) dt	\\
	&= \frac{1}{2}\int_{0}^{y}  \left(-f_{\chi^2_{n}(\frac{\rho^2}{\sigma_j^2})}(t)+f_{\chi^2_{n+2}(\frac{\rho^2}{\sigma_j^2})}(t)\right)  dt \\
	&=\int_{0}^{y}  \frac{d}{dt}f_{\chi^2_{n+2}(\frac{\rho^2}{\sigma_j^2})}(t)  dt \label{eq:der_chi_squared} \\
	&=f_{\chi^2_{n+2}(\frac{\rho^2}{\sigma_j^2})}(y),
	\end{align}
	where ${1}(\cdot)$ is the indicator function; in \eqref{eq:Poisson_representation} we used the Poisson-weighted mixture representation of the noncentral chi-square pdf; and in \eqref{eq:der_chi_squared} we used \eqref{eq:derivative_chisquared}.
\end{proof}
\begin{lemma}\label{Lem:boundedsupport}  There exists some $L=L(\sigma_1,\sigma_2,\sfR)<\infty$ such that
		\begin{align}
		& \rmN\left(\mathbb{R}, g(\cdot)+\log\left(\frac{\sigma_2}{\sigma_1}\right)-C_s\right) \notag\\
		&\quad = \rmN\left([-L,L], g(\cdot)+\log\left(\frac{\sigma_2}{\sigma_1}\right)-C_s\right)<\infty.
		\end{align}
		Furthermore, $L$ can be upper-bounded as follows:
		\begin{equation}
		L \le      \sfR d_1 +d_2 \label{eq:Bound_on_R}
		\end{equation}
		where 
		\begin{align}
		d_1&=\frac{\sigma_2+\sigma_1}{ \sigma_2-\sigma_ 1},\\
		d_2&=\sqrt{ \frac{ \frac{\sigma_2^2-\sigma_1^2}{\sigma_2^2}+2 C_s}{ \frac{1}{\sigma_1^2}-\frac{1}{\sigma_2^2} } }  \le \sqrt{ \frac{ \frac{\sigma_2^2-\sigma_1^2}{\sigma_2^2}+2 C_G}{ \frac{1}{\sigma_1^2}-\frac{1}{\sigma_2^2} } },
		\end{align}
		with 
		\begin{align} 
		 C_G(\sigma_1^2,\sigma_2^2,\sfR^2,1)  &= \frac{1}{2} \log \frac{1+\sfR^2/\sigma_1^2}{1+\sfR^2/\sigma_2^2}.
		\end{align}
	\end{lemma}

	\begin{proof}
		First,  note that $C_s \le C_G$ thanks to~\eqref{eq:C_G}.	
		Second, 	for $|y|\ge \sfR$, we can lower-bound the function $g$ as follows:
		\begin{align}
		& g(y) = \bbE\left[\log f_{Y_2^\star}(y+N) \right] - \log f_{Y_1^\star}(y) \label{eq:function_g} \\
		&= \bbE\left[\log \bbE[\phi_{\sigma_2}(y+N-X^\star) | N]  \right] - \log \bbE[\phi_{\sigma_1}(y-X^\star)] \\
		&\ge \bbE\left[\log \phi_{\sigma_2}(y+N-X^\star)   \right] - \log \bbE[\phi_{\sigma_1}(y-X^\star)] \label{eq:Jensen11} \\
		&\ge \log\frac{\sigma_1}{\sigma_2}- \bbE\left[\frac{(y+N-X^\star)^2}{2\sigma_2^2}   \right] +  \frac{(|y|-\sfR)^2}{2\sigma_1^2} \label{eq:monotonicity} \\
		&=\log\frac{\sigma_1}{\sigma_2}- \bbE\left[\frac{(y-X^\star)^2}{2\sigma_2^2}   \right]-\frac{\sigma_2^2-\sigma_1^2}{2\sigma_2^2} +  \frac{(|y|-\sfR)^2}{2\sigma_1^2} \\
		&\ge \log\frac{\sigma_1}{\sigma_2}- \frac{(|y|+\sfR)^2}{2\sigma_2^2}   -\frac{\sigma_2^2-\sigma_1^2}{2\sigma_2^2} +  \frac{(|y|-\sfR)^2}{2\sigma_1^2}, \label{eq:maxdistance}
		\end{align}
		where~\eqref{eq:Jensen11} follows from applying Jensen's inequality and the law of iterated expectation to the first term; \eqref{eq:monotonicity} follows from
		\begin{equation}
		\bbE[\phi_{\sigma_1}(y-X^\star)] \le \phi_{\sigma_1}(|y|-\sfR), \qquad |y|\ge \sfR;
		\end{equation}
		and~\eqref{eq:maxdistance} follows from $(y-X^\star)^2 \le (|y|+\sfR)^2$ for all $|y|\ge \sfR\ge |X^\star|$. The RHS of 
		\begin{align}
		&g(y)+\log\left(\frac{\sigma_2}{\sigma_1}\right)-C_s \notag\\
		&\ge  - \frac{(|y|+\sfR)^2}{2\sigma_2^2}   -\frac{\sigma_2^2-\sigma_1^2}{2\sigma_2^2} +  \frac{(|y|-\sfR)^2}{2\sigma_1^2}-C_s
		\end{align}
		is strictly positive when
		\begin{equation}
		|y|>  \frac{\sfR\left(\frac{1}{\sigma_1^2}+\frac{1}{\sigma_2^2}\right)+\sqrt{\frac{4\sfR^2}{\sigma_1^2 \sigma_2^2}+\left(\frac{1}{\sigma_1^2}-\frac{1}{\sigma_2^2}\right)\left(\frac{\sigma_2^2-\sigma_1^2}{\sigma_2^2}+2C_s\right)}}{\frac{1}{\sigma_1^2}-\frac{1}{\sigma_2^2}}.
		\end{equation}
		By using the bound $\sqrt{a +b} \le \sqrt{a} +\sqrt{b}$, we arrive at 
		\begin{align}
		|y| &\ge  \sfR \frac{\sigma_2+\sigma_1}{ \sigma_2-\sigma_1} +\sqrt{ \frac{ \frac{\sigma_2^2-\sigma_1^2}{\sigma_2^2}+2C_s}{ \frac{1}{\sigma_1^2}-\frac{1}{\sigma_2^2} } }.
		\end{align} 
		This concludes the proof for the bound on $L$. 
	\end{proof}

		\begin{lemma} \label{lem:moduls_upper_bound}
			Let $\breve{h}:\mathbb{C} \rightarrow\mathbb{C}$ denote the complex extension of the function $h$ in~\eqref{eq:productfy2}. Then, for $\sfB \ge \sfR$, we have that 
			\begin{align}
			&\max_{|z|\le \sfB} |\breve{h}(z)|  \le \frac{1}{\sqrt{2\pi\sigma_1^2}}  \rme^{ \frac{\sfB^2}{2\sigma_1^2} } \left( a_1 \sfB^2 +a_2 \sfB+a_3 \right)
			\end{align}
			where 
			\begin{align}
			a_1&=  \frac{ 3 \sigma_1^2}{ \sigma_2^2 \sqrt{\sigma_2^2-\sigma_1^2}}, \\
			a_2&=  \frac{ \sqrt{2} \sigma_1^2}{  \sqrt{ \sigma_2^2} \sqrt{\sigma_2^2-\sigma_1^2}} +2,\\
			a_3&=\frac{\sigma_1^2}{\sqrt{\sigma_2^2-\sigma_1^2}}  \left( \sqrt{  |\log(2\pi\sigma^2_2)|^2 +  \frac{ 24  (\sigma_2^2-\sigma_1^2)^2 }{\sigma^4_2}  + \pi^2}   \right).
			\end{align}
		\end{lemma}

		\begin{proof}
			Let us denote $z= z_R+i z_I$, where $z_R$ and $z_I$ are real numbers and $i=\sqrt{-1}$ is the imaginary unit. Then, by triangular inequality we have:
			\begin{align}
			&|\breve{h}(z)|  \notag\\
			&= \left|\frac{\sigma_1^2 f_{Y_1}(z) \bbE\left[N\log f_{Y_2}(z+N) \right]}{\sigma^2_2-\sigma^2_1}  \right. \notag\\
			& \quad   - \bbE\left[X^\star  \phi_{\sigma_1}(z-X^\star)\right] +zf_{Y_1}(z) \Big | \\
			&\le \left|f_{Y_1}(z) \right| \left(\frac{\sigma_1^2}{\sigma_2^2-\sigma_1^2} \bbE\big[|N|\cdot  |\log f_{Y_2}(z+N)| \big]+|z|  \right) \notag\\
			& \quad +\bbE\big[|X^\star| \cdot |\phi_{\sigma_1}(z-X^\star)|\big]. \label{eq:triangineqhbreve}
			\end{align}
			Next, let us upper-bound each contribution of~\eqref{eq:triangineqhbreve}. For~$|z|\le \sfB$, we have
			\begin{align}
			&\left|\log f_{Y_2}(z+n)\right|^2\notag\\
			 &= \left|\log\left|f_{Y_2}(z+n)\right| + i \arg(f_{Y_2}(z+n)) \right|^2 \\
			&=\log^2|f_{Y_2}(z+n)| + \arg^2(f_{Y_2}(z+n)) \\
			&=\log^2\left|  \bbE \left[ \phi_{\sigma_2}(z+n-X^\star)\right] \right| + \arg^2\left(\bbE \left[ \phi_{\sigma_2}(z+n-X^\star)\right]\right) \\
			&\le \log^2\left( \frac{1}{\sqrt{2\pi\sigma^2_2}} \bbE \left[ \exp\left(-\frac{(z_R+n-X^\star)^2-z_I^2}{2\sigma^2_2}\right)\right] \right) \notag\\
			&\quad + \arg^2\left(\sum_x \alpha_x \exp(i \theta_x)\right) \label{eq:triangineq} \\
			&\le \left( \frac{z_I^2}{2\sigma^2_2} - \frac{1}{2}\log(2\pi\sigma^2_2) + \log \bbE \left[ \rme^{-\frac{(z_R+n-X^\star)^2}{2\sigma^2_2} }\right] \right)^2+ \pi^2 \label{eq:lessthanB}  \\
			&\le 2 \hspace{-0.03cm} \left( \hspace{-0.03cm}\frac{z_I^2}{2\sigma^2_2} -\frac{1}{2}\log(2\pi\sigma^2_2) \right)^2+ \hspace{-0.03cm} 2 \log^2 \bbE \left[ \rme^{-\frac{(z_R+n-X^\star)^2}{2\sigma^2_2} }\right]  \hspace{-0.03cm} + \hspace{-0.03cm}\pi^2 \label{eq:Using_2(a+b)^2bound}  \\
			&\le 2 \left( \frac{z_I^2}{2\sigma^2_2} -\frac{1}{2}\log(2\pi\sigma^2_2) \right)^2+ 2 \hspace{-0.03cm} \frac{  \bbE^2 \left[(z_R+n-X^\star)^2 \right] }{4\sigma^4_2}  + \pi^2  \label{eq:Jensens+LogSquare}  \\
			&\le 2 \left(\frac{z_I^2}{2\sigma^2_2} -\frac{1}{2}\log(2\pi\sigma^2_2) \right)^2+ 2 \frac{   \left( (z_R+n)^2+\sfR^2 \right)^2}{4\sigma^4_2}  + \pi^2 \label{eq:UsingBound_and_zero_mean}  \\
			&\le  \frac{2 \sfB^2}{\sigma^2_2} + |\log(2\pi\sigma^2_2)|^2+  \frac{   8 (\sfB^4+n^4)+\sfR^4  }{\sigma^4_2}  + \pi^2  , \label{eq:Using_Modulus_Assumption} 
			\end{align}
			where step~\eqref{eq:triangineq} holds by triangular inequality; step~\eqref{eq:lessthanB} holds by noticing that
			\begin{equation}
			-\pi<\arg\left(\sum_{x \in \supp(P_{X^\star})} \alpha_x \exp(i \theta_x)\right)\le \pi,
			\end{equation}
			where $\{\alpha_x\}$ and $\{\theta_x\}$ are real numbers that depend on $x$;
			\eqref{eq:Using_2(a+b)^2bound} follows from using the bound $(a+b)^2 \le 2 (a^2+b^2)$; \eqref{eq:Jensens+LogSquare} holds because $ x\mapsto \log^2(x)$ is a decreasing function for $x<1$ and because $\bbE \left[ \rme^{-\frac{(z_R+n-X^\star)^2}{2\sigma^2_2} }\right]  \ge  \rme^{-\frac{ \bbE \left[(z_R+n-X^\star)^2 \right] }{2\sigma^2_2} }$, which follows from Jensen's inequality; \eqref{eq:UsingBound_and_zero_mean} follows from  $\bbE[X^\star]=0$ and  $\bbE[(X^\star)^2]\le \sfR^2$; and \eqref{eq:Using_Modulus_Assumption} follows from the bound $|a+b|^k \le 2^{k-1} (|a|^k+|b|^k)$ for $k \ge 1$; furthermore, given that $|z_R| \le \sfB$ and $|z_I| \le \sfB$, we arrive at the bound
			\begin{equation}
			\left( (z_R+n)^2+\sfR^2 \right)^2
			 \le 
			2  \left(  8 (\sfB^4+n^4)+\sfR^4 \right) .
			\end{equation}
			Consequently,
			\begin{align}
			&  \frac{\bbE\big[|N|\cdot  |\log f_{Y_2}(z+N)| \big] }{  \sqrt{ \sigma_2^2-\sigma_1^2} }\notag\\
			  &\le  \frac{\sqrt{ \bbE\big[|N|^2 \big ]  \bbE\big[ |\log f_{Y_2}(z+N)|^2 \big] }}{  \sqrt{ \sigma_2^2-\sigma_1^2} } \label{eq:applying_Caucy_Schwarz}\\
			  &\le \sqrt{   \frac{2 \sfB^2}{\sigma^2_2} + |\log(2\pi\sigma^2_2)|^2+ \frac{   8 (\sfB^4+\bbE[N^4])+\sfR^4  }{\sigma^4_2}  + \pi^2    } \label{Eq:Inserting_eq:Using_Modulus_Assumption} \\
			 &= \hspace{-0.1cm} \sqrt{  \hspace{-0.03cm}  \frac{2 \sfB^2}{\sigma^2_2} + \hspace{-0.03cm} |\log(2\pi\sigma^2_2)|^2+ \hspace{-0.03cm}  \frac{ 8 \sfB^4+24  (\sigma_2^2-\sigma_1^2)^2+ \sfR^4 }{\sigma^4_2}  + \pi^2   } ,\label{eq:applying_Lpbounds}
			 \end{align}
			 where \eqref{eq:applying_Caucy_Schwarz} follows from Cauchy-Schwarz inequality;  \eqref{Eq:Inserting_eq:Using_Modulus_Assumption} follows from $\bbE[N^4]=	3(\sigma_2^2-\sigma_1^2)^2$.
			Moreover, we have
			\begin{align}
			|f_{Y_1}(z)| &\le  \bbE\left[\left| \phi_{\sigma_1}(z-X^\star) \right|  \right] \\
			&=\frac{1}{\sqrt{2\pi\sigma_1^2}} \bbE\left[ \exp\left( -\frac{(z_R-X^\star)^2-z_I^2}{2\sigma_1^2} \right)   \right] \\
			&\le \frac{1}{\sqrt{2\pi\sigma_1^2}}  \exp\left( \frac{\sfB^2}{2\sigma_1^2} \right),  
			\end{align}
			and finally
			\begin{align}
			\bbE\big[|X^\star| \cdot |\phi_{\sigma_1}(z-X^\star)|\big] &\le \sfR\: \bbE\big[ |\phi_{\sigma_1}(z-X^\star)|\big] \\
			&\le \sfR \frac{1}{\sqrt{2\pi\sigma_1^2}}  \exp\left( \frac{\sfB^2}{2\sigma_1^2} \right).
			\end{align}
			Putting all contributions together, we get
			\begin{align}
			&|\breve{h}(z)|  \sqrt{2\pi\sigma_1^2}  \rme^{ - \frac{\sfB^2}{2\sigma_1^2} }  \notag\\
			 &\le 
			\frac{\sigma_1^2 \sqrt{   \frac{2 \sfB^2}{\sigma^2_2} + |\log(2\pi\sigma^2_2)|^2+  \frac{ 8 \sfB^4+24  (\sigma_2^2-\sigma_1^2)^2+ \sfR^4 }{\sigma^4_2}  + \pi^2   }}{\sqrt{\sigma_2^2-\sigma_1^2}} \notag\\
			&\quad  +\sfB+\sfR  \\
			&\le    a_1 \sfB^2 +a_2 \sfB +a_3, 
			\end{align}
			where in the last step we have used that $\sqrt{\sum_i  x_i } \le \sum_i \sqrt{x_i} $ and the fact that $\sfR \le \sfB$. 
		
		\end{proof}

		\begin{lemma}\label{lem:moduls_lower_bound}
			Let $\breve{h}:\mathbb{C} \rightarrow\mathbb{C}$ denote the complex extension of the function $h$ in~\eqref{eq:productfy2}. Then,
			for 
			\begin{align}
			\sfB \ge  \sfR  \frac{\sigma_2^2+\sigma_1^2}{ \sigma_2^2-\sigma_1^2}, \label{eq:Condition_on_B1}
			\end{align} 
			we have that 
			\begin{align}
			\max_{|z|\le \sfB} |\breve{h}(z)| \ge  \left( c_1 \sfB   - c_2 \sfR     \right) 
			\frac{ \exp\left(  -\frac{(\sfB+\sfR)^2}{2\sigma_1^2} \hspace{-0.1cm}\right) }{\sqrt{2\pi \sigma_1^2}}	>0,
			\end{align}
			where $c_1=1-\frac{\sigma_1^2 }{\sigma_2^2} $ and $c_2=1 +  \frac{\sigma_1^2 }{\sigma_2^2}$ .
		\end{lemma}

		\begin{proof}
			First, note that
			\begin{align}
			\frac{ \bbE_N \left[  \bbE[X^\star| Y_2=\sfB+N] \right]}{\sigma_2^2}-  \frac{\bbE[X^\star| Y_1=\sfB]}{\sigma_1^2}  
			\ge - \frac{\sfR}{\sigma_2^2}  -  \frac{\sfR}{\sigma_1^2}. \label{eq:Lower_Bound_on_CES1}
			\end{align} 
			Second, note that the condition in \eqref{eq:Condition_on_B1} implies that 
			\begin{align}
			0 \le 	\sfB\left(\frac{1}{\sigma_1^2}-\frac{1}{\sigma_2^2}\right)  - \frac{\sfR}{\sigma_2^2}  -  \frac{\sfR}{\sigma_1^2}. \label{eq:Positivity_from_condition}
			\end{align}
			Therefore, by using \eqref{eq:First_representation_h} together with \eqref{eq:Lower_Bound_on_CES1} and \eqref{eq:Positivity_from_condition}, we arrive at
			\begin{align}
			&\max_{|z|\le \sfB} |\breve{h}(z)| \ge \left|\breve{h}(\sfB) \right|\notag\\
			&= \hspace{-0.1cm} \left| \hspace{-0.05cm}   \frac{ \bbE \hspace{-0.05cm}\left[  \bbE[X^\star| Y_2=  \hspace{-0.05cm} \sfB+N  \hspace{-0.05cm}] \right]   \hspace{-0.05cm} -  \hspace{-0.05cm}\sfB}{\sigma_2^2}  \hspace{-0.05cm} -   \hspace{-0.05cm} \frac{\bbE[X^\star| Y_1=\sfB]-\sfB}{\sigma_1^2}  \hspace{-0.05cm} \right|  \hspace{-0.05cm}  \sigma_1^2 f_{Y_1}(\sfB)\\
			& \ge   \left(	\sfB\left(\frac{1}{\sigma_1^2}-\frac{1}{\sigma_2^2}\right)  - \frac{\sfR}{\sigma_2^2}  -  \frac{\sfR}{\sigma_1^2} \right)   \sigma_1^2 f_{Y_1}(\sfB)\\
			& \ge   \hspace{-0.05cm}  \left( \hspace{-0.05cm}	\sfB\left(\frac{1}{\sigma_1^2}-\frac{1}{\sigma_2^2} \hspace{-0.05cm}\right)  - \frac{\sfR}{\sigma_2^2}  -  \frac{\sfR}{\sigma_1^2} \right)   
			\hspace{-0.1cm} \frac{\sigma_1^2 }{\sqrt{2\pi \sigma_1^2}}	 \exp\left( \hspace{-0.1cm} -\frac{(\sfB+\sfR)^2}{2\sigma_1^2} \hspace{-0.1cm}\right),
			\end{align}
			where in last bound we have used  Jensen's inequality to arrive at
			\begin{align}
			f_{Y_1}(\sfB) &= \bbE\left[\phi_{\sigma_1}(\sfB-X^\star)\right] \\
			&= \frac{1}{\sqrt{2\pi \sigma_1^2}}	\bbE\left[ \exp\left(-\frac{(\sfB-X^\star)^2}{2\sigma_1^2}\right)\right] \\
			&\ge \frac{1}{\sqrt{2\pi \sigma_1^2}}	 \exp\left(-\frac{(\sfB+\sfR)^2}{2\sigma_1^2}\right). \label{eq:lb_max_h_2}
			\end{align}
			This concludes the proof. 
		\end{proof}
		
		\end{appendices} 

\bibliographystyle{IEEEtran}
\bibliography{refs.bib}
\end{document}